\documentclass[twocolumn,trackchanges]{aastex631}

\turnoffediting

\usepackage{amsmath}

\shorttitle{Protocluster-LAB connection}
\shortauthors{}

\graphicspath{{./}{figures/}}

\begin{document}
\newcommand{\ie}{$i.e.$,}
\newcommand{\mpc}{Mpc}
\newcommand{\Lya}{Ly$\alpha~$}
\newcommand{\msun}{M$_{\odot}$}
\newcommand{\cmtwo}{cm$^{-2}$}
\newcommand{\kms}{\,km~s$^{-1}$}      
\newcommand{\minpoint}{\mbox{$'\mskip-4.7mu.\mskip0.8mu$}}
\newcommand{\secpoint}{\mbox{$''\mskip-7.6mu.\,$}}
\newcommand{\sqdeg}{\mbox{${\rm deg}^2$}}
\newcommand{\squig}{\sim\!\!}
\newcommand{\subsun}{\mbox{$_{\twelvesy\odot}$}}
\newcommand{\et}{{\it et al.}~}
\newcommand{\er}[2]{$_{-#1}^{+#2}$}
\newcommand{\cgsflux}{ergs~cm$^{-2}$~s$^{-1}$}
\newcommand{\sdrel}{($1+\delta_{LAE}$)}
\def\h50{\, h_{50}^{-1}}
\def\hbl{km~s$^{-1}$~Mpc$^{-1}$}
\def\ltsima{$\; \buildrel < \over \sim \;$}
\def\simlt{\lower.5ex\hbox{\ltsima}}
\def\gtsima{$\; \buildrel > \over \sim \;$}
\def\simgt{\lower.5ex\hbox{\gtsima}} 
\def\arcsec{$''$}
\def\arcmin{$'$}
\newcommand{\ebv}{$E$($B-V$)}

\newcommand{\cpurple}    {red!75!green!50!blue}
\newcommand{\unitcgssb}  {erg\,s$^{-1}$\,cm$^{-2}$\,arcsec$^{-2}$}
\newcommand{\unitcgslum} {erg\,s$^{-1}$}
\newcommand{\sqarcsec}   {arcsec$^2$}

\title{ODIN: Where Do Ly$\alpha$ Blobs Live? Contextualizing Blob Environments \\within the Large-Scale Structure}
\author[0000-0002-9176-7252]{Vandana Ramakrishnan}
\affiliation{Department of Physics and Astronomy, Purdue University, 525 Northwestern Avenue, West Lafayette, IN 47907, USA}
\author{Byeongha Moon}
\affiliation{Korea Astronomy and Space Science Institute, 776 Daedeokdae-ro, Yuseong-gu, Daejeon 34055, Republic of Korea}
\author{Sang Hyeok Im}
\affiliation{Department of Physics and Astronomy, Seoul National University, 1 Gwanak-ro, Gwanak-gu, Seoul 08826, Republic of Korea} 
\author{Rameen Farooq}
\affiliation{Physics and Astronomy Department, Rutgers, The State University, Piscataway, NJ 08854}
\author[0000-0003-3004-9596]{Kyoung-Soo Lee}
\affiliation{Department of Physics and Astronomy, Purdue University, 525 Northwestern Avenue, West Lafayette, IN 47907, USA}
\author[0000-0003-1530-8713]{Eric Gawiser}
\affiliation{Physics and Astronomy Department, Rutgers, The State University, Piscataway, NJ 08854}
\author[0000-0003-3078-2763]{Yujin Yang}
\affiliation{Korea Astronomy and Space Science Institute, 776 Daedeokdae-ro, Yuseong-gu, Daejeon 34055, Republic of Korea}
\author[0000-0001-9521-6397]{Changbom Park}
\affiliation{Korea Institute for Advanced Study, 85 Hoegi-ro, Dongdaemun-gu, Seoul 02455, Republic of Korea}
\author[0000-0003-3428-7612]{Ho Seong Hwang}
\affiliation{Department of Physics and Astronomy, Seoul National University, 1 Gwanak-ro, Gwanak-gu, Seoul 08826, Republic of Korea}
\affiliation{SNU Astronomy Research Center, Seoul National University, 1 Gwanak-ro, Gwanak-gu, Seoul 08826, Republic of Korea}
\author[0000-0001-5567-1301]{Francisco Valdes}
\affiliation{National Optical Astronomy Observatory, 950 N. Cherry Avenue, Tucson, AZ 85719, USA}
\author[0000-0003-0570-785X]{Maria Celeste Artale}
\affiliation{Department of Physics and Astronomy, Purdue University, 525 Northwestern Avenue, West Lafayette, IN 47907, USA}
\affiliation{Physics and Astronomy Department Galileo Galilei, University of Padova, Vicolo dell’Osservatorio 3, I--35122, Padova, Italy}
\affiliation{INFN--Padova, Via Marzolo 8, I--35131 Padova, Italy}
\affiliation{Departamento de Ciencias Fisicas, Universidad Andres Bello, Fernandez Concha 700, Las Condes, Santiago, Chile}

\author[0000-0002-1328-0211]{Robin Ciardullo}
\affiliation{Department of Astronomy \& Astrophysics, The Pennsylvania
State University, University Park, PA 16802, USA}
\affiliation{Institute for Gravitation and the Cosmos, The Pennsylvania
State University, University Park, PA 16802, USA}
\author[0000-0002-4928-4003]{Arjun Dey}
\affiliation{NSF’s National Optical-Infrared Astronomy Research Laboratory, 950 N. Cherry Ave., Tucson, AZ 85719, USA}
\author[0000-0001-6842-2371]{Caryl Gronwall}
\affiliation{Department of Astronomy \& Astrophysics, The Pennsylvania
State University, University Park, PA 16802, USA}
\affiliation{Institute for Gravitation and the Cosmos, The Pennsylvania
State University, University Park, PA 16802, USA}
\author[0000-0002-4902-0075]{Lucia Guaita}
\affiliation{Departamento de Ciencias Fisicas, Universidad Andres Bello, Fernandez Concha 700, Las Condes, Santiago, Chile}
\author[0000-0002-2770-808X]{Woong-Seob Jeong}
\affiliation{Korea Astronomy and Space Science Institute, 776 Daedeokdae-ro, Yuseong-gu, Daejeon 34055, Republic of Korea}
\author[0000-0001-9850-9419]{Nelson Padilla}
\affiliation{Instituto de Astronomía Teórica y Experimental (IATE), Comisión Nacional de Investigaciones Científicas y Técnicas (CONICET), Universidad Nacional de Córdoba, Laprida 854, X500BGR, Córdoba, Argentina}
\author{Akriti Singh}
\affiliation{Departamento de Ciencias Fisicas, Universidad Andres Bello, Fernandez Concha 700, Las Condes, Santiago, Chile}
\author[0000-0001-6047-8469]{Ann Zabludoff}
\affiliation{Steward Observatory, University of Arizona, 933 North Cherry Avenue, Tucson AZ 85721}
\begin{abstract}
While many Ly$\alpha$ blobs (LABs) are found in and around several well-known protoclusters at high redshift, how they trace the underlying large-scale structure is still poorly understood. In this work, we utilize 5,352 Ly$\alpha$ emitters (LAEs) and 129 LABs at $z=3.1$ identified over a $\sim$ 9.5 \sqdeg\ area in early data from the ongoing One-hundred-deg$^2$ DECam Imaging in Narrowbands (ODIN) survey to investigate this question.
Using LAEs as tracers of  underlying matter distribution, we identify overdense structures as galaxy groups, protoclusters, and filaments of the cosmic web.
We find that LABs preferentially reside in regions of higher-than-average density and are located in closer proximity to overdense structures, which represent the sites of protoclusters and their substructures.  Moreover, protoclusters hosting one or more LABs tend to have a higher descendant mass than those which do not. Blobs are also strongly associated with filaments of the cosmic web, with $\sim$ 70\% of the population being within a projected distance of $\sim$ 2.4 pMpc from a filament. 
We show that the proximity of LABs to protoclusters is naturally explained by their association with filaments as large cosmic structures are where many filaments converge. 
{The contiguous wide-field coverage of the ODIN survey allows us to firmly establish a connection between LABs \emph{as a population} and filaments of the cosmic web for the first time.}

\end{abstract}


\section{Introduction} \label{sec:intro}

In the local universe, galaxies in overdense environments tend to be more massive \citep[e.g.,][]{van_der_Burg2013} and are more likely to be quiescent \citep[e.g.,][]{Peng2010,Quadri2012}. At $z$ $\gtrsim$ 1.5, this trend weakens \citep{Alberts2014,Alberts2016,Nantais2016,Kawinwanichakij2017} or even reverses \citep{Elbaz2007,Hwang2019,Lemaux2022}. At $z\gtrsim2 $, the highest-density regions -- believed to be sites of the progenitors of present-day  galaxy clusters, or protoclusters -- display copious star formation and AGN activity, often in excess of that observed in regions of average density \citep[e.g.,][]{Casey2015,Umehata2015,Oteo2018,Harikane2019,Shi2020}. To gain insight into how large-scale environment influences the evolution of galaxies  over cosmic time, it is necessary to study a large sample of overdense structures at high redshift and the galaxy inhabitants therein. 

Lacking many of the observable markers of fully virialized clusters, protoclusters are often identified as overdensities of galaxies such as dusty star-forming galaxies \citep{Oteo2018}, Lyman break galaxies \citep[e.g.,][]{Toshikawa2016,Toshikawa2018}, H$\alpha$ emitters \citep[e.g.,][]{Hayashi2012,Darvish2020,Koyama2021}, or \Lya emitters \citep[e.g.,][]{Lee2014,Jiang2018,Harikane2019,Higuchi2019}. Alternatively, several `signposts' have been explored as promising avenues to find them. These include radio galaxies and QSOs, and more recently,  \Lya nebulae, referred to as Lyman alpha blobs \citep[LABs: see][for a review]{Overzier2016}.

LABs are extended luminous Ly$\alpha$ sources, $L_{{\rm Ly}\alpha}$$\sim$~$10^{43}$--$10^{44}$~\unitcgslum\ and $\ge50$~kpc in size \citep{Francis1996, Steidel2000, Dey2005, Yang2009, Yang2010, Ouchi2020}. While what powers their emission remains poorly constrained, possible mechanisms include galactic super-winds \citep{Taniguchi2000}, ionizing photons from star formation \citep{Geach2016, Ao2017} and AGN activity \citep{Dey2005, Geach2009, Yang2014a, Cai2017}, resonant scattering of \Lya photons \citep{Hayes2011, You2017, Kim2020,Chang2022}, and gravitational cooling \citep{Fardal2001, Rosdahl2012, Daddi2021, Fabrizio22}.  
LABs often host multiple galaxies and are sometimes associated with overdense regions \citep{Steidel2000, Matsuda2004, Prescott2008, Matsuda11, Yang2010, Badescu2017,Kikuta2019} or cosmic filaments \citep[e.g.,][]{Erb2011,Umehata2019}, providing a promising pathway to study  protoclusters.   

How LABs are distributed within the large-scale structure remains unclear. Some studies find tentative evidence that the morphologies of LABs are aligned with the large-scale structure \citep[e.g.,]{Erb2011,Kikuta2019} and that the brightest blobs tend to lie near the densest regions \citep[e.g.][]{Kikuta2019}. Meanwhile \citet{Badescu2017} observed that LABs appear to avoid the most overdense regions and to prefer the outskirts of massive structures. Other studies find that not all LABs reside in overdense environments \citep[e.g.,][]{Hibon2020}. Many of these results are based on a single protocluster and/or a small sample of LABs, making it difficult to properly account for the effect of cosmic variance and to address the question of how reliably LABs trace protocluster sites. 
To make significant progress, it is essential to study the relationship between LABs and their large-scale environment \textit{in a statistical manner}.

One efficient way to achieve this goal is by conducting a narrow-band imaging survey aimed at finding both LABs and the more compact and commonplace \Lya emitting galaxies (LAEs). LAEs are generally young, low-mass star-forming galaxies \citep{Gawiser2006,Gawiser2007,Guaita2011} whose relatively low galaxy bias ($b\approx 2$) and low luminosity imply that they trace the bulk of the high-redshift galaxy population \citep{Gawiser2007}, making them ideal tracers of  the large-scale structure \citep[e.g.,][]{huang22}. Simultaneously, \Lya emission at $z$ $\gtrsim$ 2 is redshifted into the visible window, facilitating detection over large areas from the ground using wide-field imagers. 

In this work, we utilize the early science data from the ongoing  One-hundred-deg$^2$ DECam Imaging in Narrowbands (ODIN) survey, the widest-field narrowband survey to date. The large sample of LAEs (5,352) selected over a wide (9.5~deg$^2$) contiguous field allows us to peer into a well-defined slice of the cosmos in which groups, filaments, and voids are readily visible. Equipped with this information, we investigate where 129 LABs at $z=3.1$ live in the context of the large-scale structure spanning hundreds of comoving Mpcs. Through this work, we hope to demonstrate the power of wide-field narrow-band imaging in illuminating cosmic structure formation in ways that cannot be easily replaced by the upcoming ELTs.


This paper is organized as follows. In Sections~\ref{sec:data} and \ref{sec:sample_selection}, we describe the imaging data and the selection of the LAE and LAB samples, respectively. In Section \ref{sec:lae_ods}, we explore multiple methods to map the large-scale structure using LAEs as tracers. 
%
We  examine the relationship between  LABs with the measured large-scale environment in Section \ref{sec:analysis} and summarize our  findings in Section \ref{sec:summary}.
Throughout this paper, we adopt a cosmology with $\Omega=0.27$, $\Omega_{\Lambda}=0.73$, $H_0=70$ km~s$^{-1}$~Mpc$^{-1}$ \citep{Komatsu2011}. Distance scales are given in comoving units of $h_{70}^{-1}$ cMpc, with the h$_{70}$ suppressed unless noted otherwise. All magnitudes are given in the AB system \citep{Oke1983}.

\section{Data and Catalogs} \label{sec:data}

\subsection{ODIN and SSP Imaging Data} \label{subsec:img_data}

\begin{figure}
    \centering
    \includegraphics[width=1.1\linewidth]{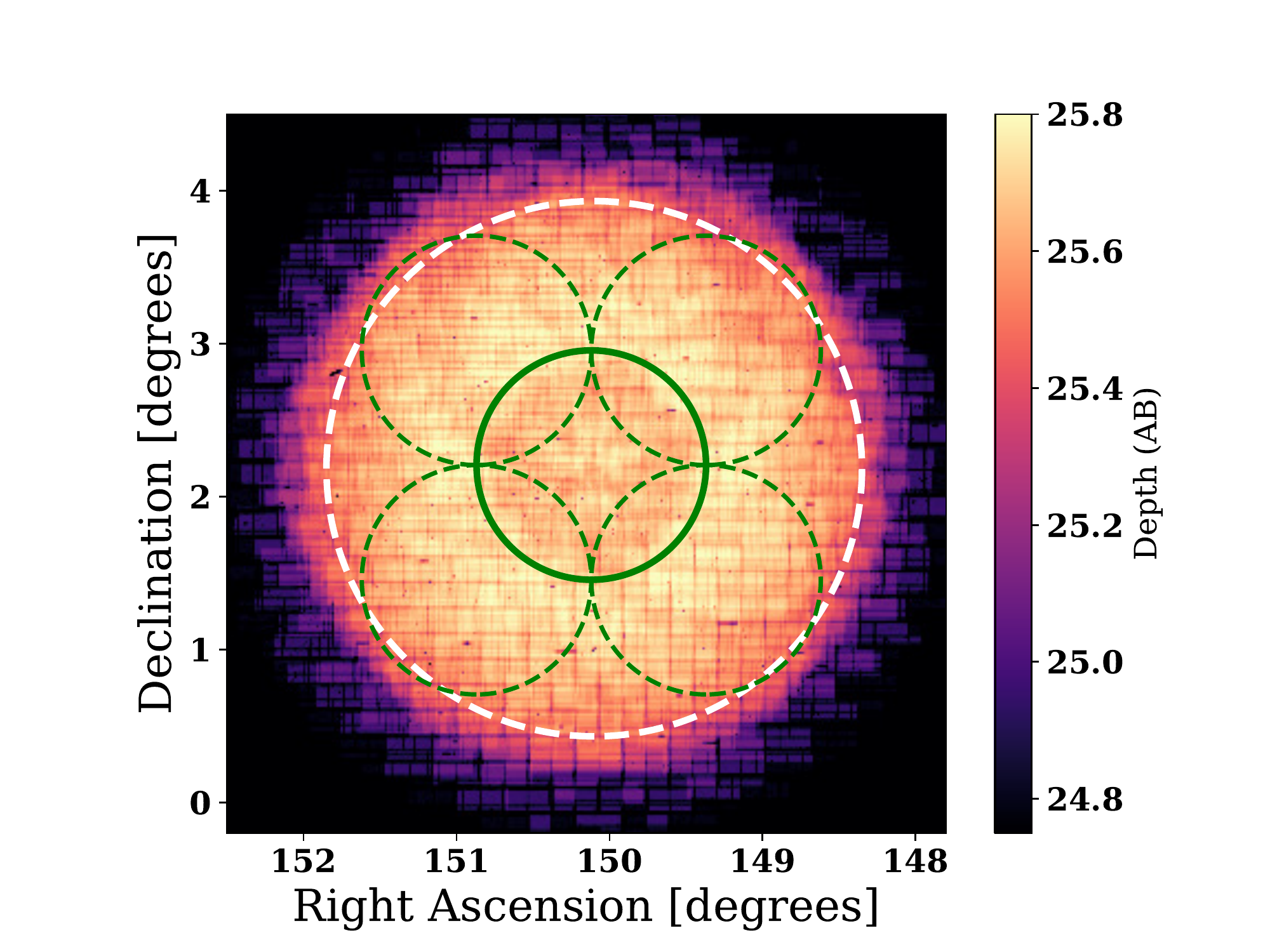}
    \caption{The $5\sigma$ depth of the ODIN E-COSMOS $N501$ data  is indicated by the colorbar on right. The white dashed line indicates the anticipated coverage of the LSST Deep Drilling Field. Green circles mark the positions of the SSP Deep (dashed) and UltraDeep (solid) pointings.
    }
    \label{fig:data_depths}
\end{figure}

As a survey program approved by the NSF's Optical-Infrared Laboratory, ODIN is currently undertaking the widest narrow-band imaging survey to date using the Dark Energy Camera \citep[DECam,][]{Flaugher2015} on the Blanco Telescope at the Cerro Tololo Inter-American Observatory. 
Using three custom narrow-band (NB) filters ($N419$, $N501$, and $N673$ filters), ODIN is covering seven contiguous fields totaling 91~deg$^2$ in area, each sampled at three redshifts, $z=2.4$, 3.1, and 4.5. The details of the survey design, data reduction, and calibration will be presented in a separate paper (K.-S. Lee et al. in preparation). In this work, we analyze a single ODIN field  observed with our $N501$ filter.  
The filter characteristics ($\lambda_C/\Delta \lambda$=5015/73 \AA)  are  sensitive to the redshifted \Lya emission at $z$ $\sim$ 3.1 (3.09 $<$ $z$ $<$ 3.15). The data covers $\sim$ 12 \sqdeg\ of the extended COSMOS field\footnote{Five of the ODIN fields are designed to match the LSST field of view of its Deep Drilling Fields and their pointing centers. For the COSMOS field, it is $\alpha$=10:00:24, $\delta$=02:10:55 (J2000).} with seeing 0.9\arcsec\ at a near-uniform depth of 25.6~mag 
in the central 10 deg$^2$. 
The depth and coverage of the $N501$ data are shown in Figure \ref{fig:data_depths}.


We  make use of the publicly available $grizy$ broad-band (BB) data from the HyperSuprimeCam Subaru Strategic Program \citep[SSP:][]{Aihara2018a,Aihara2018b} from the second data release \citep{Aihara2019}. The sky area mapped by the  SSP survey is smaller than the ODIN coverage, limiting the area in which LAEs can be selected to $\sim$ 9.5 \sqdeg. In Table~\ref{tab:data_depths}, we list the 5$\sigma$ limiting magnitudes of all bands. These are based on the 5$\sigma$ fluctuation of the noise measured in randomly placed 2\arcsec\ diameter apertures.  The SSP coverage of the COSMOS field consists of one  UltraDeep pointing and four Deep pointings \citep{Aihara2018a} as shown as green solid and dashed circles, respectively, in Figure~\ref{fig:data_depths}.  As a result,  the variation of the BB imaging depths is significant. The effect of the depth variation on the detection of LAEs is discussed in Section \ref{subsec:laes}. 

\begin{deluxetable}{ccc}
\tablecaption{\label{tab:data_depths}}
\tablecolumns{3}
\tablehead{ \colhead{Band} & \colhead{Depth (Deep/UltraDeep)} & \colhead{Seeing} }
     \startdata
         $N501$ & 25.6/25.6 & 0.90\arcsec \\
         $g$ & 26.3/26.6 & 0.81\arcsec \\
         $r$ & 26.0/26.2 & 0.74\arcsec \\
         $i$ & 25.9/26.0 & 0.62\arcsec \\
         $z$ & 25.8/26.0 & 0.63\arcsec \\
         $y$ & 24.8/25.2 & 0.71\arcsec \\
    \enddata
    \tablecomments{Median depth and seeing of the imaging data. The depth is measured as the 5$\sigma$ fluctuation of the noise in 2\arcsec\ diameter apertures. 
    }
\end{deluxetable}



The $N501$ data consists of 72 individual DECam exposures (each with exposure time of 1200~s)  taken in February 2021; the total  observing time is 24~hrs for the field with a per pixel exposure time range of 20~min to 7.3~hrs and average of 2.9~hrs or 3.1~hrs (minimum of 1 or 2 overlapping exposures respectively). Individual DECam frames are processed and coadded with the DECam Community Pipeline \citep{Valdes2014,DCP} into a single image.

Each of the 62 DECam CCDs is flat-fielded separately by dome flats, star flats, and dark sky illumination flats.  Master dome flats are produced by stacking sequences of 11 exposures taken nightly. Star flats are produced periodically from 22 widely dithered exposures of a field with many bright stars using the algorithm of  \citet{2017PASP..129k4502B}. Dark sky illumination flats are created by coadding unregistered stacks of exposures.

The background is measured in blocks by the modes of non-source pixels. The background is then made uniform by matching the means in each CCD and subtracting a low-order fit to the modes. While this step is critical to producing a uniform dithered stack, it leads to over-subtraction of the faint halos around bright stars.  However, we remove any science source close to bright stars in the analysis by applying star masks (Section~\ref{subsec:sources}). Thus, the effect of uneven background levels near bright stars on the small, distant extra-galactic objects is negligible. 

An astrometric solution is derived for each CCD by matching stars to Gaia-EDR3 \citep{2021A&A...649A...1G}. The higher order distortions are predetermined and fixed, and the low order terms are updated using the astrometric solver SCAMP \citep{2006ASPC..351..112B} with continuity constraints between CCDs. The solution RMS is typically 10s of milliarcseconds. The solution is used to reproject the exposures to a standard tangent plane sampling with constant pixel sizes using sinc interpolation. A fixed tangent point for all the exposures in the field is used. The exposures are matched to the Pan-STARRS-1 photometric catalog \citep{ 2012ApJ...756..158S} for a flux zero point to provide the scaling and, along with seeing and sky brightness estimates, weighting of the coadd. The dithered exposures are stacked by averaging registered pixels with statistical rejection (constrained sigma clipping) of outliers to minimize cosmic rays accumulated from the long exposures.
 
Following the format of the SSP data release, the final ODIN stack is split into multiple `tracts', each $1.7^\circ \times 1.7^\circ$ in size with an overlap of 1\arcmin. The SSP tracts are 
reprojected using the DECam Community Pipeline to have the same tangent points and pixel scales (0.26\arcsec) as the ODIN data. 

\subsection{Source detection} \label{subsec:sources}

Source detection is conducted using the \textsc{Source Extractor} software \citep{Bertin1996} run in dual image mode using the $N501$ band data as the detection image while performing photometry in all bands. For PSF-matched photometry, rather than degrading the images with smoothing kernels, we measure the flux in successive, closely spaced apertures. The appropriate aperture correction for a given band is  computed by requiring that the fraction of the flux enclosed remains constant. Regardless, for the aperture size we choose for LAE  selection (2\arcsec\ diameter) the correction  is minimal for all filters. 
Assuming a Moffat profile with $\beta$ = 2.5, the aperture correction factor for a point source varies from 1.07 to 1.09 when seeing changes from  0\farcs6 to 1\farcs0. 


A great majority of LAEs are expected to be point sources  at $z=3$ \citep{Malhotra2012,Paulino_Afonso2018}. Prior to source detection, all images are convolved with a Gaussian filter ({\tt FILTERING}=Y) with a full-width-at-half-maximum (FWHM) matched to the seeing value of the $N501$ data, optimizing for the detection of point sources \citep{Gawiser2006b}.  The detection threshold in the filtered image is set to 0.95$\sigma$  and the minimum area is set to 1 pixel ({\tt DETECT\_THRESH} = 0.95 and {\tt DETECT\_MINAREA} = 1). The choice of {\tt DETECT\_THRESH} is motivated by running \textsc{Source Extractor} on sky-subtracted and inverted (or `negative') versions of our science images. In these negative images, any detected sources are due to noise fluctuations, as all the true sources will have pixel values well below zero. If the noise is Gaussian, the fluctuations of the sky value both above and below the mean should be the same; thus, the number of sources detected in the negative images should represent the extent of the contamination of the source catalog by noise peaks. To maximize the detection of faint sources, we choose the minimum value of {\tt DETECT\_THRESH} that yields a contamination fraction of less than 1\%.    
 We remove objects with a signal-to-noise ratio (S/N) less than 5 in a 2\arcsec\ diameter aperture ($N501 \gtrsim 25.6$) and the sources with internal flags {\tt FLAG}$\geq$4 -- suggesting that they contain saturated pixels or significant image artifacts -- from our catalog. We also use the  star masks released as part of the SSP DR2 \citep{Coupon2018} and remove all sources  near bright stars. Accounting for the sky area excluded by these masks, the effective area covered by our catalog is $\sim$ 7.5 \sqdeg, and the total number of $N501$-detected objects after making these cuts is 689,962.

\subsection{Simulations}

While a detailed comparison of our results with the expectations from state-of-the-art hydrodynamic simulations is beyond the scope of this work, we make use of the IllustrisTNG simulations here to build cosmologically sound expectations for how cosmic structures may manifest themselves in observations such as ODIN.  To this end, we use  the IllustrisTNG300-1 (hereafter TNG300) simulation, the largest box with the highest resolution available from the IllustrisTNG suite \citep[][]{Nelson2019,Pillepich2018a,Pillepich2018b}. TNG300 represents a periodic box of 302.6~cMpc on a side and is run from $z=127$.  
The cosmological parameters for the TNG simulation are  different from ours\footnote{The IllustrisTNG simulations adopt the Planck cosmology \citep{Planck2016_cosmology}: $\Omega_{\rm \Lambda} = 0.6911$, $\Omega_{\rm b} = 0.0486$, $\Omega_{\rm m} = 0.3089$, $H_{0} = 100\,h\,{\rm km}\,s^{-1}\,{\rm Mpc}^{-1}$ with $h=0.6774$}. 

In addition to the publicly available TNG data, we also make use of the UV magnitudes computed by \citet[][]{Vogelsberger2020}. A Ly$\alpha$ luminosity is assigned to each halo following the prescription given in \citet{Dijkstra2012,Weinberger2019}. Both UV and Ly$\alpha$ luminosity functions computed within the full TNG300 volume are in good agreement with the measurements in the literature. A full description of the procedures and the predictions for protocluster galaxy populations will be presented in M.C. Artale et al. (in preparation). 

\section{Sample Selection} \label{sec:sample_selection}

\subsection{Lyman-$\alpha$ Emitter selection} \label{subsec:laes}

\begin{figure*}[t]
    \centering
    \includegraphics[width=0.80\linewidth]{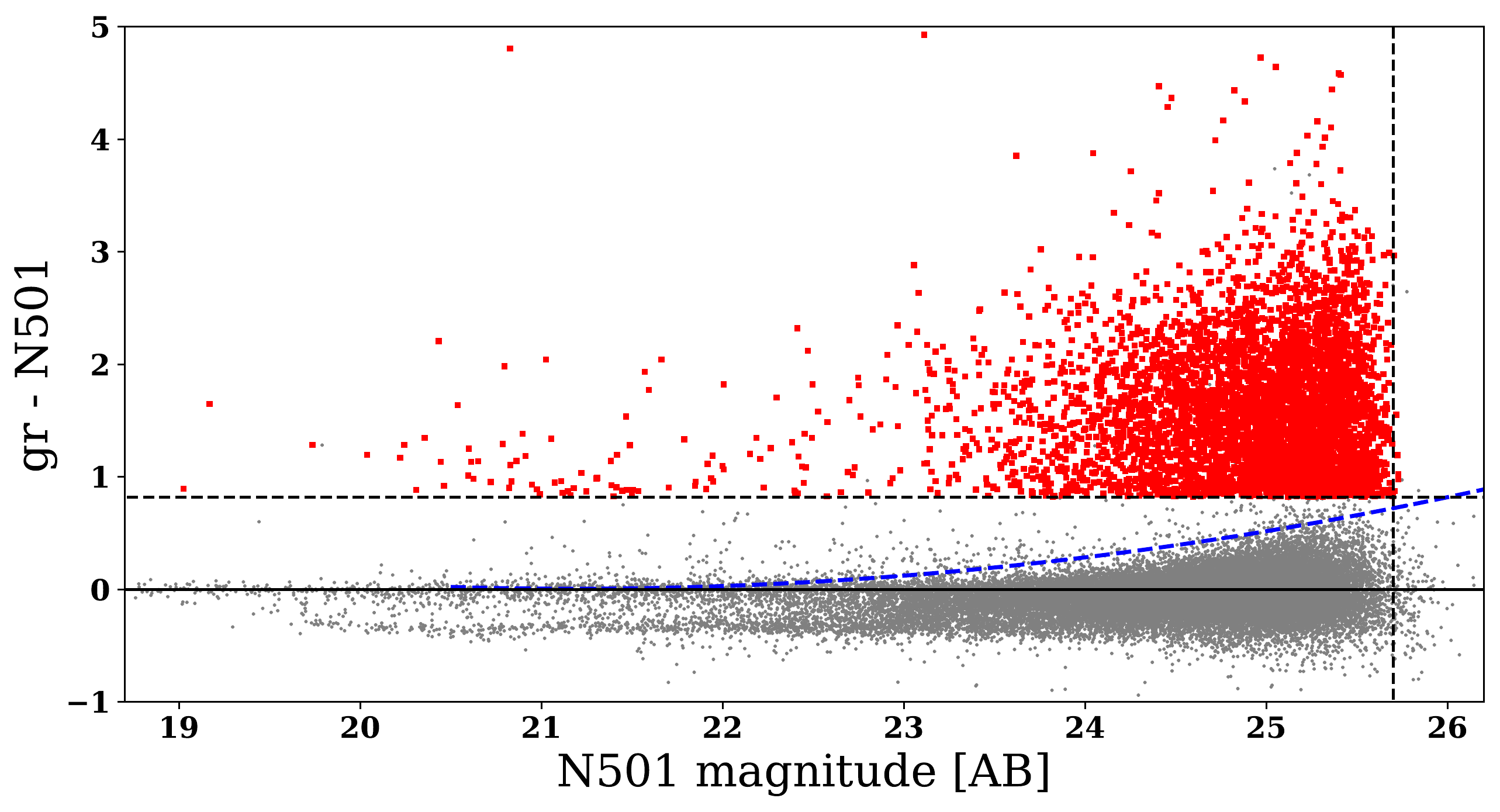}
    \caption{Narrow-to-broad-bnd color excess ($gr$ - $N501$) vs $N501$ magnitude. 
    Red points represent LAE candidates. Grey dots show 1 in every 20 $N501$-detected sources that are not LAEs. Black  dashed lines mark the color cut corresponding to $W_0=20$~\AA\ (horizontal) and 5$\sigma$ limiting magnitude of the $N501$ data (vertical), respectively. The blue dashed line shows the median 3$\sigma_{gr-N501}$ line as a function of $N501$ magnitude. 
    }
    \label{fig:lae_sn}
\end{figure*}
The details of ODIN LAE selection methods will be presented in a forthcoming paper (N. Firestone et al., in preparation) and we only briefly summarize it here. We select LAEs as sources with an NB excess, based on the NB-continuum color. \citet{Gronwall2007} and \citet{Gawiser2007} found that  $z\sim3$ LAE samples selected via narrowband excess corresponding to rest-frame equivalent width $W_0>20$~\AA\ suffer greater contamination from continuum-only objects than from [O~{\sc ii}] emitters.  In order to obtain a robust estimate of the 501 nm continuum level of all objects in the catalog, we create a weighted average of the $g$ and $r$ band flux density in a 2\arcsec~diameter aperture, using weights  determined from the central wavelengths of $g$, $r$, and $N501$ to estimate the flux density at 501~nm:
\begin{equation}  f_{gr} \equiv 0.83f_g + 0.17f_r   \;   \;    \end{equation}
We convert $f_{gr}$ to an AB magnitude  $gr$ and  select all objects with color excess $gr-N501>0.82$, which corresponds to  $W_0>20$ {\AA} following the equation: 
\begin{equation}  (gr-N501) > 
2.5 \log \left ( 1 + \frac{[\lambda_\mathrm{eff}/\lambda_{{\rm Ly}\alpha,0}]\,W_0}
{\Delta \lambda_{N501}}\right )   \;   \;     \end{equation}
where $\Delta \lambda_{N501}$ is the FWHM  of the $N501$ filter transmission (72.5~\AA), $\lambda_{{\rm Ly}\alpha,0}$ is the rest-frame wavelength of Ly$\alpha$ (1215.67~\AA), and 
$\lambda_\mathrm{eff}$ is the observed-frame Ly$\alpha$ wavelength, i.e., the central wavelength of $N501$ (5015~\AA).  
Additionally, we remove all objects whose NB color excess is consistent with zero at the $3\sigma$ level to minimize contamination from continuum-only objects scatter into the color cut by requiring: 
\begin{equation}
gr-N501 > 3 \sigma_{gr-N501}    \; \;  
\end{equation}
The photometric scatter in the $gr-N501$ color,  $\sigma_{gr-N501}$, is calculated by propagating the uncertainties of the flux densities in each band. {The median color scatter is well fit by a second-order polynomial of the form, $\sigma_{gr-N501}=5.2259 - 0.4934m + 0.01165m^2$ where $m$ is the $N501$ magnitude.} Our selection results in 5,352 LAE candidates in our sample. In Figure~\ref{fig:lae_sn}, we plot the $gr - N501$ color versus the $N501$ magnitude for the full catalog along with the selected LAE candidates. {The blue dashed line shows the functional form of the median  3$\sigma_{gr-N501}$ line given above.} Our chosen $gr - N501$ cut places our LAE candidates safely above the locus of continuum-only objects.

While we leave a detailed  assessment of our LAE candidates for future work, the number of LAE candidates we find is in reasonable agreement with the expectations from previous studies. \citet{Gronwall2007} found 162 $z$ $=$ 3.1 LAEs in 0.28~{\sqdeg} with fluxes above 1.5 $\times$ 10$^{-17}$ {\cgsflux} with a 50~{\AA}  filter; \citet{Gawiser2007} reported a 20\% uncertainty in the resulting LAE number density once cosmic variance due to large-scale structure  is included. \citet{Ciardullo2012} found 130 $z$ $=$ 3.1 LAEs in 0.28~{\sqdeg} with fluxes above 2.4 $\times$ 10$^{-17}$ \cgsflux\ with a 57~{\AA}  filter. Accounting for the width of our filter and conservatively assuming Poissonian error, we would expect $\sim$ 6549 $\pm$ 515 LAEs based on the result of \citet{Gronwall2007} and 4887 $\pm$ 429 LAEs based on the result of \citet{Ciardullo2012} in our 7.5~\sqdeg\ survey area. 
{Given differences in completeness and depth across various surveys, it is difficult to concretely quantify the contamination fraction for our sample based on these past works, but the good agreement of the observed LAE number density with the expectations suggests that the contamination fraction is low. A spectroscopic program to confirm a portion of our LAE candidates is ongoing, and the results of this program together with a proper estimation of the contamination fraction will be presented in a future work. We also note that since the contamination fraction should be more or less constant across the entire field, the presence of contaminants will not have a significant effect on the estimation of the LAE surface density, which is the main purpose of the LAE sample in the present work.}

The depth variation of the SSP BB data across our survey field  (see Table~\ref{tab:data_depths}) should not significantly affect the LAE number density. While greater BB depth would in principle reduce the uncertainty of the estimated $gr$ magnitude, the uncertainty on the $gr - N501$ color excess is dominated by the photometric scatter in the $N501$ band, which has a uniform coverage. This can be seen in Figure~\ref{fig:lae_sn}, where the median 3$\sigma_{gr-N501}$ value is very small for sources with bright $N501$ magnitudes and increases rapidly with increasing $N501$ magnitude. Indeed, we find that the LAE number density in the UltraDeep region (790~deg$^{-2}$) is consistent with that in the Deep region (756~deg$^{-2}$) within the Poisson uncertainty.

\begin{figure}
    \includegraphics[width=\linewidth]{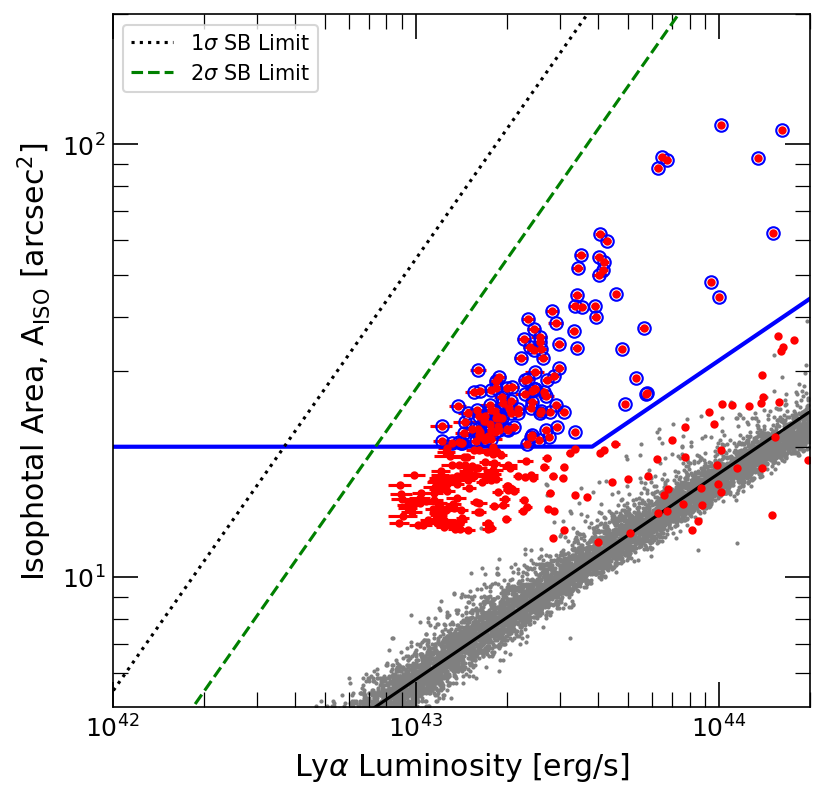}
    \caption{
    Positions of our LAB candidates are shown as large red circles highlighted in blue on the $A_{\mathrm{iso}}$-$L_{{\rm Ly}\alpha}$ space. Small red circles mark all {the sources with $A_{\mathrm{iso}}>3$ \sqarcsec} while
    grey dots and the thick solid line indicate the locations  of simulated point sources and the best-fit scaling law, respectively. Blue solid lines outline the final LAB selection criteria: (1) $A_{\mathrm{iso}}>20$~{\sqarcsec}; and (2) sources lie $>3\sigma$ above the relation of point sources. 
    The diagonal dotted and dashed lines correspond to the 1$\sigma$ and 2$\sigma$ surface brightness limits, respectively. 
    \label{fig:LAB_selection}}
\end{figure}

\begin{figure*}
    \centering
    \includegraphics[width=1\textwidth]{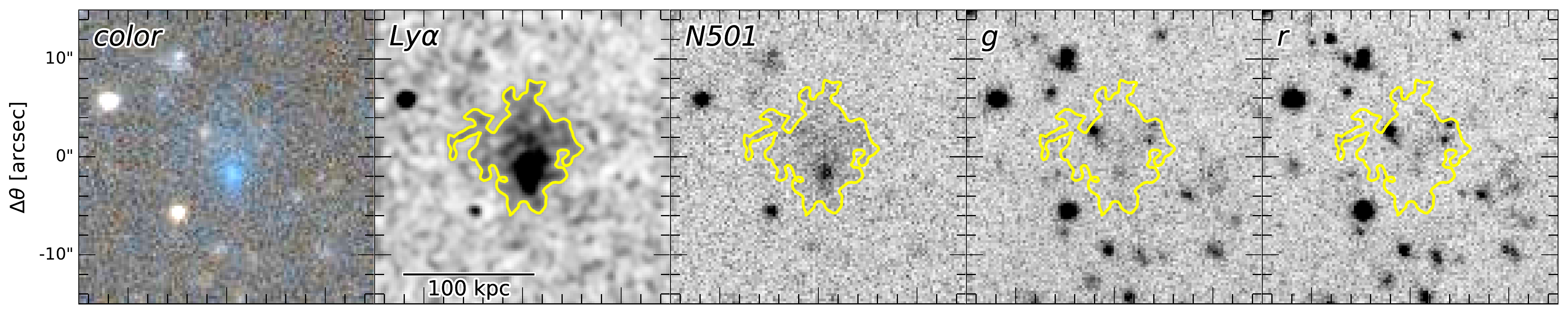}
    \caption{
    An ODIN LAB at $z=3.1$ with spectroscopic confirmation. The postage stamp images are  30\arcsec\ on a side. The color image (leftmost) is created with the DECaLS $rg$ and $N501$ data used as RGB, respectively. The BB and NB images are from the SSP and ODIN $N501$ data. In the four right panels, yellow contours outline our SB threshold, $3.3\times10^{-18}$~\unitcgssb.  Multiple galaxies are found within or near the Ly$\alpha$-emitting region, which extends nearly $\approx$100~kpc.
    }
    \label{fig:LAB_example}
\end{figure*}

\subsection{Ly$\alpha$ Blob Selection} \label{subsec:labs}
The details of the final selection of ODIN LABs will be presented in another  paper (B. Moon et al., in preparation). Here, we provide a brief description. Our selection method is similar to those used in previous blind LAB searches \citep[e.g.,][]{Matsuda2004, Yang2010}.
To look for extended \Lya emission, we select LABs in two steps:
(1) identifying objects with narrow-to-broad-band color excess (i.e., as LAEs) and 
(2) detecting extended \Lya emission around them from a continuum-subtracted Ly$\alpha$ image.

To detect the bright core of LABs, we first create another LAE catalog using detection settings and color criteria that are slightly different  than those given in Sections~\ref{subsec:sources} and~\ref{subsec:laes}. We choose a higher {\tt DETECT\_THRESH} of 1.2$\sigma$ and a larger {\tt DETECT\_MINAREA} = 4 to exclude faint sources that are associated spurious low surface brightness features.
Then, we apply the following criteria: 
(1) $N501 < 25.62$ and 
(2) $gr - N501 > 0.8$, 
where all fluxes and magnitudes are measured in a 2\arcsec\ diameter aperture. 

To detect extended \Lya emission, we create a  \Lya  image by subtracting the continuum flux from the $N501$ image, with the continuum flux estimated from the $g$ and $r$ bands as described in Section \ref{subsec:laes}. {The broad-band images are smoothed with a Gaussian kernel to match their PSF sizes with narrow-band image before the continuum subtraction.}
We generate a mask  for areas with negative sky counts and halos around saturated stars in the $gr$ bands, which can mimic diffuse emission. The mask is used as {\tt MAP\_WEIGHT} to prevent the detection of such features. After filtering the image with a 7-pixel 2D Gaussian filter with FWHM of 3 pixels, we detect all sources with contiguous isophotal size greater than 42 pixels ($\sim$3 \sqarcsec) all of which rise above the surface brightness threshold of $3.3\times10^{-18}$~\unitcgssb. The value corresponds to 1.5$\sigma$ where $\sigma$ is the pixelwise sky rms measured in the \Lya image. 
From {the sources with isophotal size over 3 \sqarcsec}, we select those with an isophotal area greater than 20~arcsec$^2$. We further require that at least one LAE conicide with the extended emission for the source to be considered an LAB candidate.


In Figure~\ref{fig:LAB_selection}, we show as red circles the distribution of isophotal  sizes ($A_{\mathrm{iso}}$) and $L_{{\rm Ly}\alpha}$ of all recovered sources. Grey dots represent similar measurements made for simulated point sources. To guard against bright point sources being selected as LABs, we require that LAB candidates lie above the $3\sigma$  line of the known $A_{\rm iso}-L_{{\rm Ly}\alpha}$ relation for point sources and that $A_{\mathrm{iso}}\geq 20$ \sqarcsec. {This minimum area threshold is of the same order as those adopted by previous narrowband surveys for LABs \citep[e.g.]{Matsuda2004, Yang2009, Yang2010}. The precise LAB selection criteria were determined after extensive tests, varying the minimum area and the surface brightness threshold while visually inspecting the resulting LAB candidates. These tests suggest that most of the additional LAB candidates selected by relaxing $A_{\mathrm{iso}}$ to a lower value \citep[such as 16 \sqarcsec, similar to][]{Matsuda2004} are spurious. A quantitative comparison of various LAB selection criteria will be provided in a forthcoming paper (B. Moon et al, in preparation).}These criteria are indicated by blue lines in Figure~\ref{fig:LAB_selection}. A total of 129  LAB candidates are identified in our final sample; these are shown in Figure~\ref{fig:LAB_selection} as large red circles highlighted in blue. 

Given the difference in sensitivity of various surveys, differences in selection criteria, and strong field-to-field variations \citep{Yang2010}, it is difficult to directly compare the number density of our LAB candidates with those found in existing surveys. As the ODIN survey progresses further, we will robustly quantify these variations based on the LAB statistics from seven widely separated fields at a uniform depth. 

\begin{figure}
    \includegraphics[width=\linewidth]{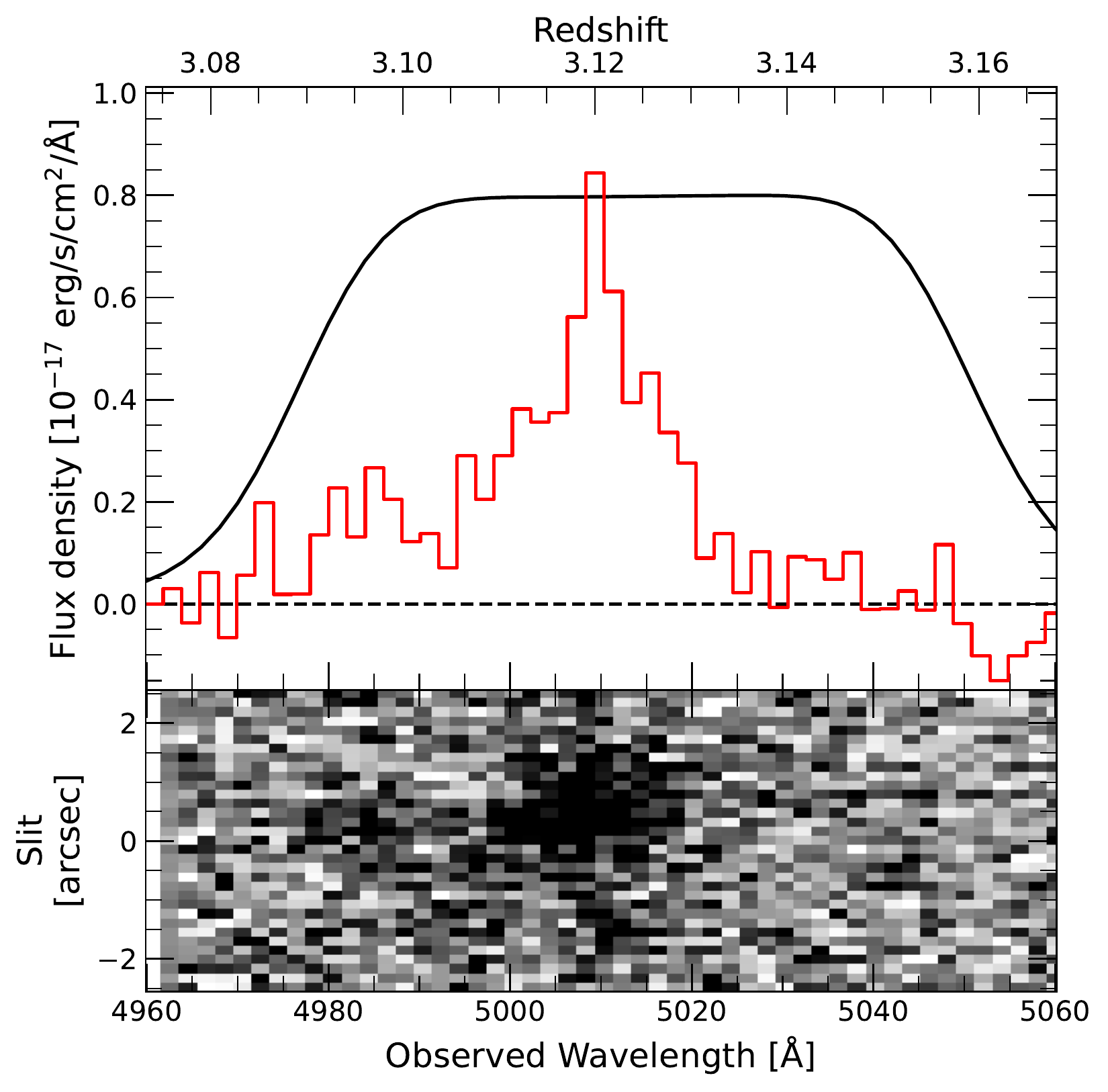}
    \caption{
    {1D and 2D spectrum from the LAB shown in Figure~\ref{fig:LAB_example}. Gemini/GMOS reveals spatially extended and diffuse \Lya emission in 2D spectrum. The 1D spectrum, extracted from the 2D spectrum, shows a double-peaked \Lya profile.}
    The black line represents the transmission curve of $N501$ in arbitrary units. 
    }
    \label{fig:LAB_confirm}
\end{figure}

Figure~\ref{fig:LAB_example} shows one of our LAB candidates in $grN501$ bands as well as the Ly$\alpha$ image. The source has $L_{{\rm Ly}\alpha}= 6.5\times10^{43}$~\unitcgslum\ and  $A_{\rm iso}$ = 96 \sqarcsec. Several galaxies lie within or near  the Ly$\alpha$-emitting region, which extends to $\approx$100~kpc. These characteristics are similar to Ly$\alpha $ blobs discovered in the past \citep[e.g.,][]{Steidel2000, Matsuda2004, Yang2011, Yang2014b, Prescott2012}. 

In Spring 2022, several LAB candidates were targeted by a Gemini/GMOS program and subsequently confirmed, which includes the LAB shown in Figure~\ref{fig:LAB_example}. Followup of more LAB targets is scheduled in 2023. While the full results of the spectroscopic programs will be presented elsewhere, Figure~\ref{fig:LAB_confirm} shows the 1D \Lya spectrum for the LAB. The black line indicates the $N501$ transmission normalized arbitrarily. {The profile shows a very wide width ($sim$ 1400 km/s, FWHM) and a double peaked line profile with a weaker blue peak, characteristic of \Lya emission. The absence of any other emission lines (e.g., H$\alpha$, H$\beta$, [OIII]) makes it very unlikely that the source is an [OII] or [OIII] emitter at lower redshift. Furthermore, it is unlikely for lines other than \Lya to display the observed extended emission, since they do not undergo resonant scattering.}  In addition to the spectroscopically confirmed LABs, our selection recovers RO-0959, a known LAB at $z=3.096$  published by \citet{Daddi2021}, even though its line emission falls on the edge of the $N501$ transmission. Confirmation of these LABs lends support to the robustness of our LAB selection.




\section{Tracing the large-scale structure traced with LAEs} \label{sec:lae_ods}
Galaxies are biased tracers of the underlying matter distribution. Thus, once the galaxy bias of a given population is known, their positions can be used to map the large-scale structure. Generally, existing studies have found that more massive or more luminous galaxies tend to have higher galaxy biases than their less luminous cousins as they occur preferentially in the high-density peaks \citep{Kaiser1984,Davis1985,Norberg2002}. 
 Existing studies also suggest that LAEs have the lowest bias value of all probed galaxy populations at high redshift \citep[$\sim$ 2,][]{Gawiser2007,Guaita2010,Khostovan2019, Hong2019}. Their high abundance and low bias make them excellent tracers of the underlying matter distribution \citep[see, e.g.,][]{huang22}. Here, we study the large-scale structure at $z$ $\sim$ 3.1 traced by the LAEs in our sample. {One caveat, however, is that LAEs, particularly those in the most overdense regions, could be hidden by dense clouds of HI gas \citep[e.g.][]{Shimakawa2017,Momose2021}. This could cause us to underestimate the surface density towards the cores of protoclusters.} 


\begin{figure}
    \centering
    \includegraphics[width=\linewidth]{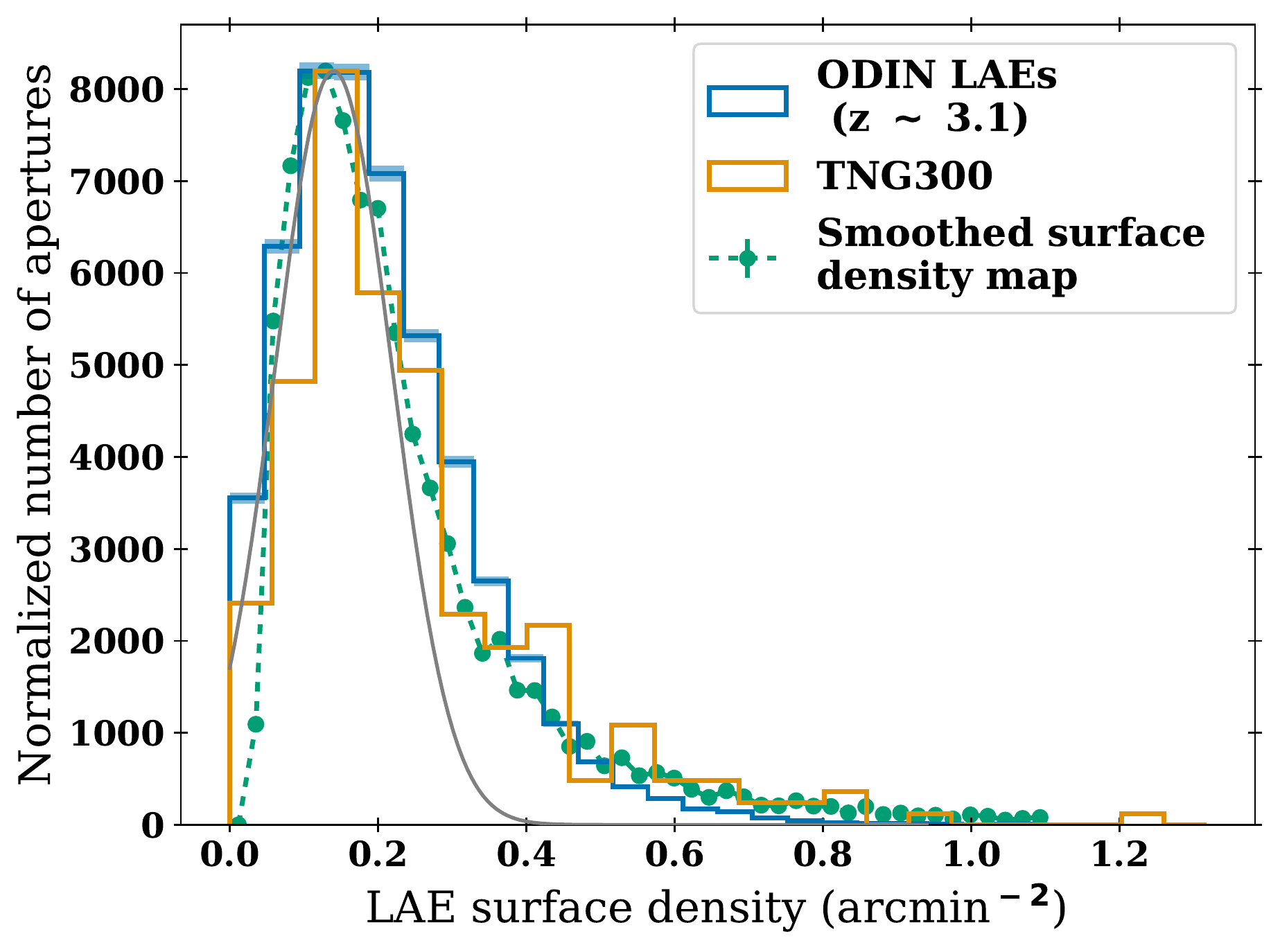}
    \caption{The distribution of LAE surface densities  measured within randomly distributed 5~cMpc radius circular apertures is shown in blue. Similar measurements made on the TNG300 simulations (orange) with a line-of-sight thickness that matches the NB width are in reasonable agreement with our data.  The green histogram shows the LAE density map constructed by smoothing the LAE positions with a Gaussian kernel (Section~\ref{subsec:gauss_smoothing}). All three  exhibit a clear excess at the high-density end over {a Gaussian approximation of the Poissonian function} expected for a purely random distribution. In both data and simulations, the highest LAE overdensity regions trace the largest cosmic structures. 
    }
    \label{fig:sd_in_apers}
\end{figure}

In Figure \ref{fig:sd_in_apers}, we show the distribution of the LAE surface density across the field by placing a 5~cMpc (2.6\arcmin) radius circle on 20,000 randomly chosen positions and measuring the number of LAEs enclosed therein. If LAEs show no clustering, they are expected to obey a Poisson distribution  as shown by a dashed line. The fact that the distribution shows a significant excess at high surface densities strongly suggests that they are in fact clustered. 

We repeat the measurements using the LAEs modeled in the TNG300 simulations. The line-of-sight `thickness' of our data determined by the $N501$ filter transmission is matched by carrying out the measurements on a randomly chosen 300 $\times$ 300 $\times$ 60~cMpc cosmic volume sliced along the X, Y, or Z direction of the simulation. The results are shown in orange. While the simulated LAE counts slightly overpredict at the high-density end, the overall distributions of the real and simulated LAEs are qualitatively similar with a well-matched peak occurring at $\approx 0.2$~arcmin$^{-2}$. The significant excess of the regions of high LAE densities seen in both data and simulations suggests the presence of large cosmic structures. 

In this section, we explore different ways to use LAEs as tracers of cosmic structures thereby detecting groups, protoclusters, and filaments of the cosmic webs.  

\begin{figure*}
    \centering
    \gridline{\fig{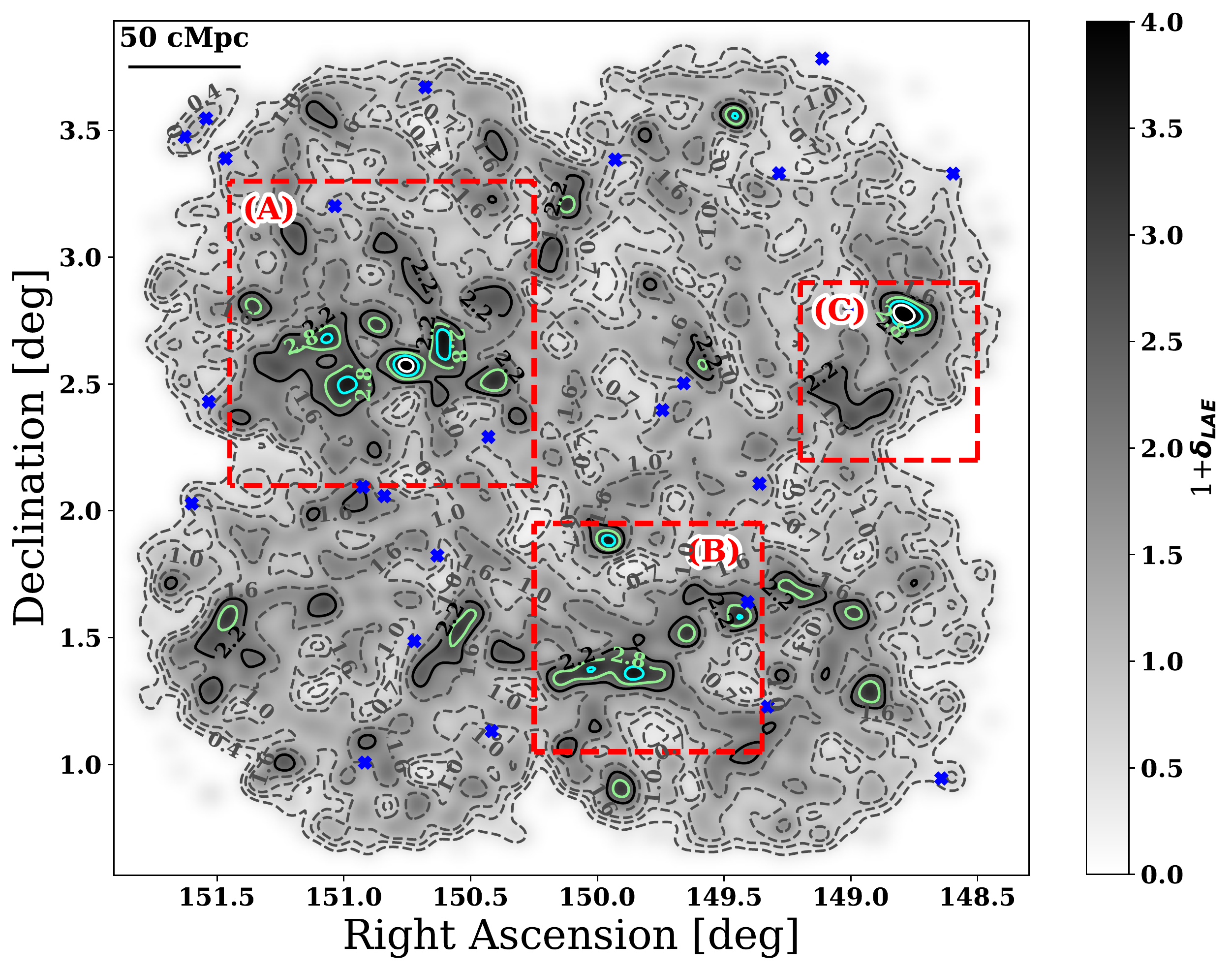}{\linewidth}{}}
    \vspace{-1.5em}
    \gridline{\fig{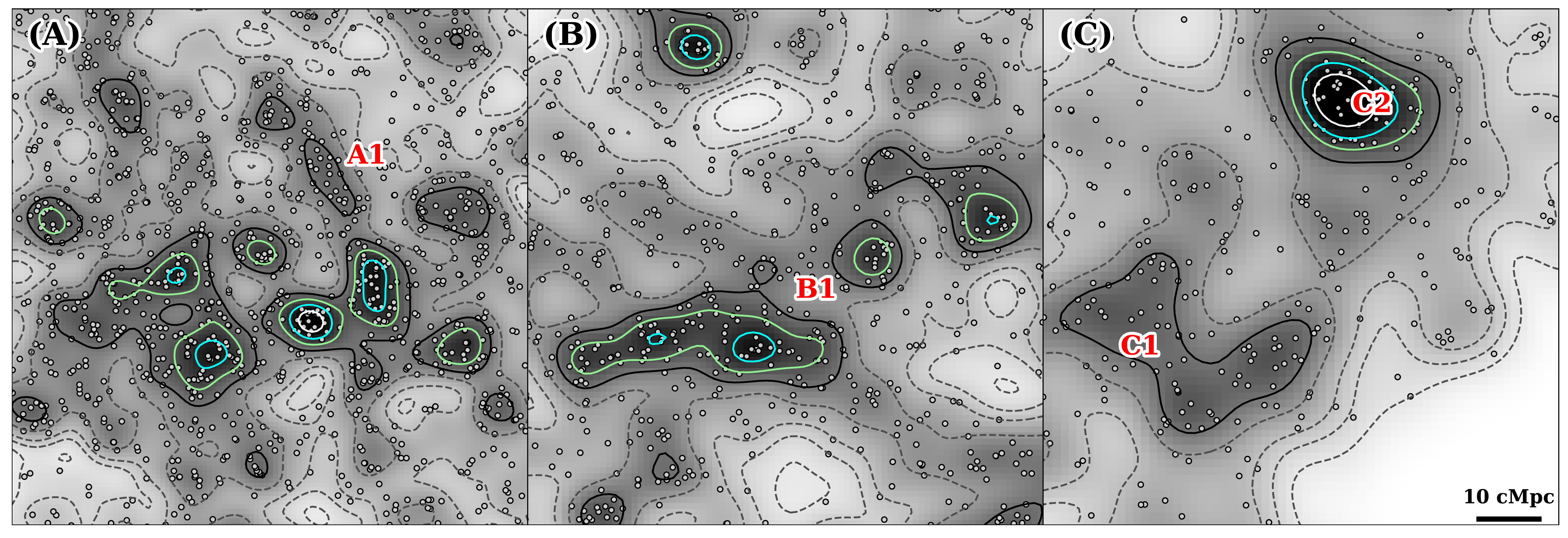}{\linewidth}{}}
    \caption{\emph{Top:}  LAE overdensity map constructed from the Gaussian kernel smoothing method (Section~\ref{subsec:gauss_smoothing}) is shown in both greyscale and contour lines. 
     Black, green, blue, and white contours indicate overdensities with $\delta_{LAE}$ 2-, 3-, 4- and 5$\sigma$ above the field value, respectively. {The position of the holes left by bright stars, which are filled in with uniformly distributed random points, are indicated by blue crosses.} \emph{Bottom:} Zoomed-in views of the three regions outlined by dashed rectangles. Individual LAEs are indicated as gray dots. Some features of interest are labeled.
    }
    \label{fig:lae_sd_maps_gauss}
\end{figure*}

\subsection{Gaussian kernel smoothed density map} \label{subsec:gauss_smoothing}
The simplest and the most commonly used method of creating a surface density map is by smoothing the LAE distribution with a fixed kernel \citep[e.g.,][]{Yang2010,Lee2014,Saito2015,Badescu2017,Shi2019,Zheng2021,huang22}. In addition to being straightforward to implement, it produces a visualization that is easy to understand. 

We begin by creating an LAE number density map with a pixel size of 0.01$^\circ$ (1.15~cMpc at $z=3.1$). The empty regions left by bright stars and image defects are filled in by populating {uniformly distributed random points that match the mean density of the field. A uniform random distribution is a resaonable approximation of the LAE distribution since, as discussed in Section \ref{subsec:laes}, the LAE surface density is determined predominantly by the narrowband depth, which is highly uniform in our data (see Figure \ref{fig:data_depths}.} The map is then convolved with a Gaussian kernel whose size is determined using Kernel Density Estimation (KDE) following the method given in  \citet{Badescu2017}. 
%

In KDE, an estimator $\hat{f}(x)$ is created for the  underlying distribution $f(x)$ 
from which a set of data points arise by smoothing the data with a predetermined kernel. 
The best kernel size, referred to as bandwidth in KDE, is determined via the leave-one-out cross validation scheme as follows. 
The estimator $\hat{f}_{-i}(x;\sigma)$ is found using a Gaussian kernel with width $\sigma$ and leaving out the $i^{\rm th}$ data point $x_i$. The likelihood of the estimator yielding the $i^{\rm th}$ data point is  $\hat{f}_{-i}(x_i;\sigma)$. The $\sigma$ value 
that optimizes the likelihood of finding all data points is the one that
maximizes
$\prod_{i}\hat{f}_{-i}(x_i)$. For our LAE sample,  the optimal  Gaussian kernel has FWHM = 5\arcmin.2 (10~cMpc at $z=3.1$).  
This kernel size is comparable to the expected size of a protocluster \citep{Chiang2013}. The distribution of the LAE surface density using this method is shown in Figure~\ref{fig:sd_in_apers} (green), consistent with other measurements therein.
      
The LAE (surface) overdensity is computed as: 
\begin{equation}
    \delta_{LAE}~=\frac{~\Sigma_{\rm LAE}}{\overline{\Sigma}_{\rm LAE}} - 1
    \label{eq:lae_od}
\end{equation}
where $\overline{\Sigma}_{\rm LAE}$ and $\Sigma_{\rm LAE}$ are the mean and local LAE density, respectively. {The mean density and its standard deviation are determined by fitting the $\Sigma_{\rm LAE}$ histogram to a Gaussian function of the form $\exp{[-(N-\mu)^2/2\sigma^2]}$. The fitting is done after sigma-clipping the high tail with a threshold of 1.5$\sigma$, retaining only the low end of the distribution which is representative of the field for the fitting. A Gaussian function is expected to be a good fit to the low end of the surface density distribution \citep[e.g.][]{Chiang2014,ArayaAraya2021}; this is also seen in Figure \ref{fig:sd_in_apers}. We find $\mu=0.14$~arcmin$^{-2}$ and $\sigma=0.08$~arcmin$^{-2}$, respectively}.

In the top panel of Figure~\ref{fig:lae_sd_maps_gauss}, we show the relative LAE density, \sdrel. The contours indicate overdensities at the $2\sigma$ (black), 3$\sigma$ (green), $4\sigma$ (blue), and $5\sigma$ (white) levels. {In the forthcoming discussion, we refer to LAE overdensities of 2-3$\sigma$ as `moderate' densities and LAE overdensities of $>$ 3$\sigma$ as `high' densities} Multiple large complexes of overdensities are visible within which several hundreds of LAEs reside. Three of the largest complexes, labeled as A, B, and C in the figure, are shown in the bottom panels of Figure~\ref{fig:lae_sd_maps_gauss} where individual LAE positions are indicated. The morphology of these LAE overdensities  is strikingly irregular, which we summarize below: \\

\noindent {\it - Complex A:} the largest structure in our map -- has at least four individual groups. In addition, an elongated moderate-density structure (labelled A1)  extends northeast from the largest group. The configuration is reminiscent of a filamentary arm connected to a massive halo. \\

\noindent {\it - Complex B:} Similar to A, multiple regions of overdensities are connected by `bridges' (one of them labelled B1) of more moderate overdensity. B1 is not captured well in this smoothed density map. This topic will be revisited in Section~\ref{subsec:voronoi}.\\

\noindent {\it - Complex C:}  An extended structure (C1) is connected to a more overdense one (C2) via a filament. Once again, the filament is not evident from the contour lines but can be seen from the alignment of LAEs stretching out from C2 soutward. \\

The features seen in these complexes -- such as elongated structures, clumpy morphology, and filaments connecting large structures -- are similar to those seen in cosmological dark matter simulations \citep[e.g.,][]{Bolyan_Kolchin2009,Kuchner2022} and are in qualitative agreement with  expectations from the hierarchical theory of structure formation. 

While the Gaussian kernel smoothing method does an excellent job of pinpointing significant overdensities, it does not fare well in detecting intermediate-density features such as filaments. This shortcoming is tied to the choice of the smoothing scale (10~cMpc), which is applied in all directions. Any structure whose size is comparable to or larger than this value would stand out clearly in the smoothed map whereas those smaller or narrower than this scale would not. To circumvent this challenge, we take a scale-free approach in Section~\ref{subsec:voronoi}.

\begin{figure*}
    \gridline{\fig{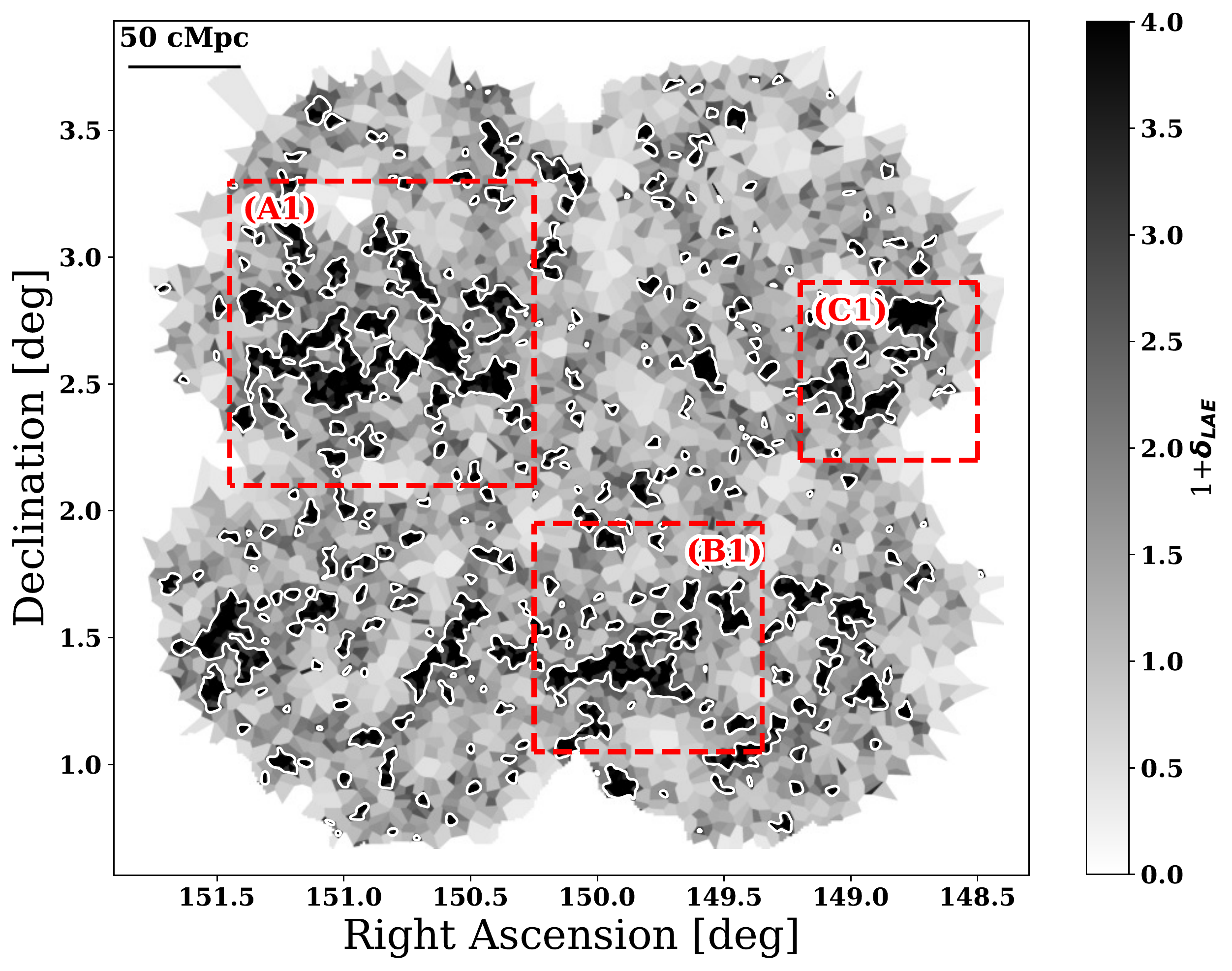}{\linewidth}{}}
    \vspace{-1.5em}
    \gridline{\fig{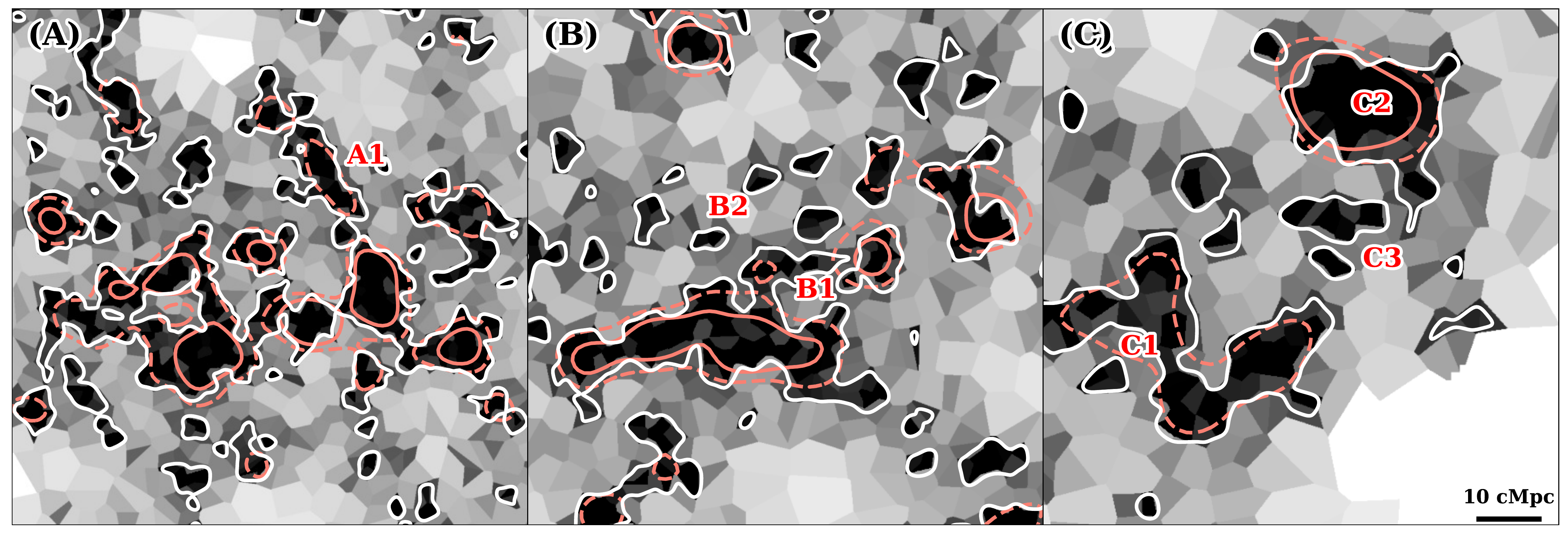}{\linewidth}{}}
\caption{
\emph{Top:} LAE overdensity map constructed from the Voronoi tessellation (Section~\ref{subsec:voronoi}). White lines indicate 3$\sigma$ overdensity contours. The bottom panels show the  three regions -- labeled A, B, and C -- in the top panel. Overlaid in pink are the $2\sigma$ (dashed) and $3\sigma$ (solid) contour lines from the GS map.  While the GS and VT methods recover similar structures, the latter fares better in detecting anisotropic/intermediate-density structures than the former. Several features of interest are labeled and discussed in text.
}
\label{fig:lae_sd_maps_voronoi}
\end{figure*}

\subsection{Voronoi tessellation} \label{subsec:voronoi}
Tessellation-based methods perform well at finding small-scale and/or anisotropic structures \citep{Darvish2015} and have been employed in several recent studies \citep[e.g.,][]{Dey2016,Lemaux2018,Cucciati2018,Hung2020,Malavasi2021}. Here, we apply the Voronoi tessellation (VT) method to the LAE positions.

VT takes the locations of a set of points and  partitions the space occupied by them into cells. Each cell is constructed to contain one generating point (in this case, a galaxy) and is comprised of the points that are closer to the enclosed generating point than any other. The size of a Voronoi cell is taken as a measure of the density of the surrounding region, i.e., cells that fall in an overdense (underdense) region will be smaller (larger) in area than that of an average LAE. 

We estimate the LAE surface density as follows. First, we calculate the area of the Voronoi cells, $A_V$, corresponding to each LAE. Any cell  larger than 0.01 \sqdeg\ ($\sim$ 130~cMpc$^2$ at $z=3.1$) is excluded from further analysis as such cells are unphysically large. Visual inspection confirms that these cells are  unbounded and  are located at the edges of the image. Such cells comprise $\approx$ 2\% of the total number. The surface density of an LAE is  the inverse of the area of the Voronoi cell in which it is located. 

Following the prescription given in the literature \citep[e.g.,][]{Cucciati2018,Lemaux2018,Hung2020}, we construct a pixellated density map based on VT by populating the field with a uniform grid of points; the spacing of the grid is 3\farcs6 (0.12~cMpc), much smaller than the Voronoi polygons. All points within a given polygon are assigned the same density ($=1/A_{V}$). 

Similar to the GS map, the mean density is determined by fitting the density histogram with a Gaussian function. We find that the best-fit parameters are ($\mu$,$\sigma$)=(0.10,0.06)~arcmin$^{-2}$, in reasonable agreement with those determined in Section~\ref{subsec:gauss_smoothing}. The overdensity  map is generated using Equation~\ref{eq:lae_od}. 

The resultant map is shown in the top panel of Figure~\ref{fig:lae_sd_maps_voronoi} and reveals the same structures discussed in Section \ref{subsec:gauss_smoothing}. As expected, the tessellation method fares better in detecting smaller and more irregular structures. In this context, we reexamine Complexes A, B, and C. To facilitate comparison, we display the GS map $2\sigma$ and $3\sigma$ contours in pink.  \\

\noindent{\it - Complex A:} The  filamentary arm-like structure labeled as A1 is  more clearly detected in the VT map compared to the GS map (Figure~\ref{fig:lae_sd_maps_gauss}, {bottom left}). It is detected at the same significance as the galaxy group to which it is connected. \\

\noindent {\it - Complex B:} 
Similarly to A, the `bridge' labeled B1 connecting the largest elongated structure to a smaller one in the northwest is clearly detected at a high significance. This feature is not fully captured in the GS map. Several smaller overdensities in the region labelled B2 are newly detected in the VT map.\\

\noindent {\it - Complex C:}  the irregular overdensity in the southeast (C1) is clearly delineated with a higher significance than previously. The region labeled C3 connecting the two largest overdensities (C1 and C2) is newly detected in the VT map.  \\

All in all, many of the most significant structures have clumpy/irregular morphologies consisting of multiple closely clustered overdensities, which are often joined together by bridges of moderate density. These features are in qualitative agreement with the expectations from the hierarchical theory of structure formation and affirm the notion that LAEs do trace the underlying large-scale structure of matter distribution, including the cosmic web. 

\subsection{Detection of Cosmic Structures} \label{subsec:finding_struct}

 In this section, we describe how we identify cosmic structures at $z=3.1$ traced by LAEs, by using the density maps discussed in Sections~\ref{subsec:gauss_smoothing} and \ref{subsec:voronoi}. We refer to these LAE-overdense structures as protoclusters, used in a broad sense. First, from the GS map, we define overdense structures as regions enclosed by the 3$\sigma$  contours corresponding  to \sdrel\ $=$ 2.84 with a minimum area of 78.5~cMpc$^2$ ($\sim$22~arcmin$^2$). The motivation for the latter requirement is to ensure that the projected size of a detected structure is at least as large as that of the adopted kernel ($\pi \cdot 5^2=78.5$). While the condition would be easily satisfied by any protocluster\footnote{A protocluster at $z=3$ that will evolve into a galaxy cluster with masses $M_{z=0}\geq 10^{14}~h_{100}^{-1}M_\odot$ has the half-mass radius of 5--10~cMpc \citep{Chiang2013}}, it may exclude smaller groups unless they are close to the main halo.

\begin{figure*}
    \centering
     \includegraphics[width=6.5in]{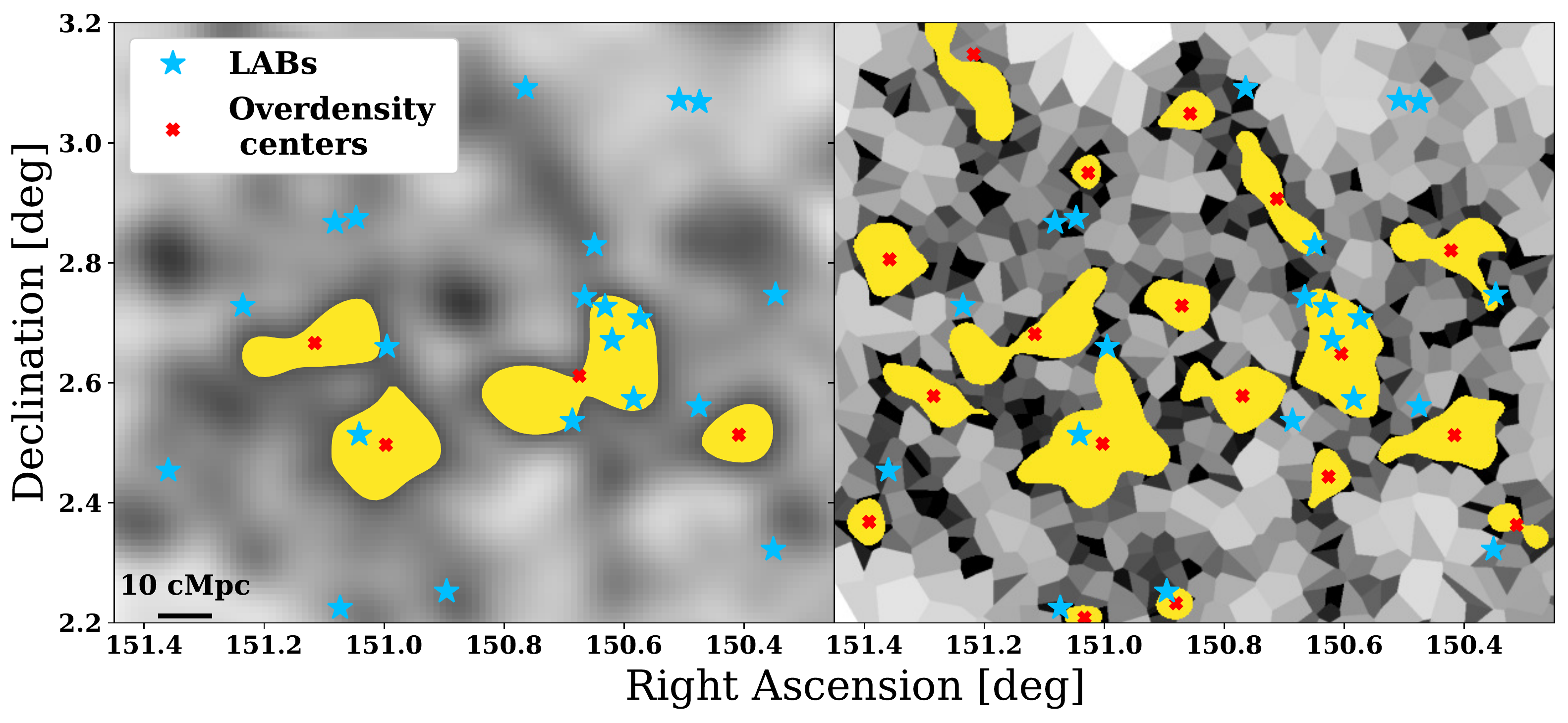}
    \caption{Structures detected in Complex A by the GS (left) and VT (right) density maps are shown as yellow swaths. The geometric centers of the structures are marked by red crosses.    While both methods identify the most significant structures with a similar angular extent, the VT map fares better in detecting smaller and/or elongated structures and in deblending structures in close proximity. The locations of Ly$\alpha$ blobs (blue stars) relative to the detected structures hint at the possible correlation. 
    } 
    \label{fig:detected_ods}
\end{figure*}

For each structure, we assign the geometric center of the  contour as its center. In most cases, the center location does not change significantly even if we define it as the  peak density region  instead. Of the 12 
structures we detect, three have coordinates offset by more than 2.6\arcmin\ ($\sim$5~cMpc). These are located at ($\alpha$,$\delta$)=(149.9$^\circ$,1.4$^\circ$), (150.7$^\circ$,2.6$^\circ$), and (151.1$^\circ$,2.7$^\circ$). These protoclusters are  irregular/elongated in their morphology; for example, the protocluster at (150.7$^\circ$,2.6$^\circ$) is clearly a blend of two systems (as seen in Figure~\ref{fig:detected_ods}) and should perhaps not be treated as one. The optimization of protocluster/group selection will be presented in future work. At present, we note that even using the density peak as the center of the structures instead of the geometric center, our results remain qualitatively unchanged.

We also utilize the VT map to detect structures adopting a procedure similar to those in the literature \citep[e.g.,][]{Lemaux2018,Hung2020,Sarron2021}. We use \textsc{SEP} \citep{Barbary2016}, a Python implementation of the \textsc{SourceExtractor} software. Prior to detection, we internally  smooth the VT map with a Gaussian kernel with FWHM 5~cMpc. Doing so helps to minimize the number of false detections and obtain relatively smooth boundaries for a given structure. As expected, the number of detections depends sensitively on the threshold  and the minimum area. We adopt  {\tt DETECT\_THRESH} = 4$\sigma${, corresponding to an absolute LAE overdensity value of 2.4,} and {\tt DETECT\_MINAREA} = 7.7 arcmin$^2$ (25 cMpc$^2$). Once again, the latter is comparable to the smoothing scale, i.e., detected structures are always larger. Since the smoothing kernel is smaller than the required minimum area, the shape of the detected overdensities is not significantly affected by the smoothing. This can be seen in the right panel of Figure~\ref{fig:detected_ods} where the shape of the regions selected as overdensities is reasonably well preserved. {The former is chosen based on the result of \citep{Chiang2013}, who find that protoclusters with descendant mass $\gtrsim$ 3 $\times$ 10$^{14}$ M$_\odot$ should display narrowband-selected LAE overdensities of $\sim$ 1.5 - 2.5. We leave a detailed discussion the protocluster selection parameters to a forthcoming paper (Ramakrishnan et al, in prep.)}

In Figure~\ref{fig:detected_ods}, we illustrate the extent of the detected structures and their centers in Complex A identified from the GS and the VT maps. While the five largest overdensities are detected by both with similar sizes, the VT method fares better at picking up smaller and/or more irregular overdensities. For example, the filament extending northeast (labeled A1 in Figures~\ref{fig:lae_sd_maps_gauss} and \ref{fig:lae_sd_maps_voronoi}) is only detected in the VT map. It also performs better at separating nearby structures. The pair of overdensities around (150.6$^\circ$, 2.6$^\circ$) is identified in the GS map as one structure. Overall, we find that the VT method is preferable for structure detection to fixed kernel smoothing. However, we make use of both sets of structures in the subsequent analysis to demonstrate the robustness of our results against the specifics of structure detection.

\subsection{Cosmic Filaments  traced by LAEs} \label{subsec:filaments}

Our visual examination reveals many linear features connecting extended  structures traced by LAEs. Similar features have been observed by spectroscopic surveys in galaxy positions around massive protoclusters \citep[e.g.][]{Cucciati2018,huang22}. Motivated by this, we identify cosmic filaments based on the LAE positions using the Discrete Persistent Structure Extractor code  \citep[DisPerSE:][]{Sousbie2011a}. Given a set of  coordinates, DisPerSE constructs a density map based on the Delaunay tessellation, and identifies local maxima, minima, and saddle points using the Hessian matrix. Starting from each saddle point, it creates a small segment that runs parallel to the eigenvector of the matrix with a positive eigenvalue. From the end of this segment, the next segment is computed that runs parallel to the gradient vector of the density field. This procedure is repeated until the segments reach a local maximum. Finally, the collection of these segments is extracted as filaments. More details are  provided in \citet{Sousbie2011a, Sousbie2011b} while details of the parameters used in this work are given in Appendix~\ref{appendix_B}. 

 \begin{figure*}
    \centering
    \gridline{\fig{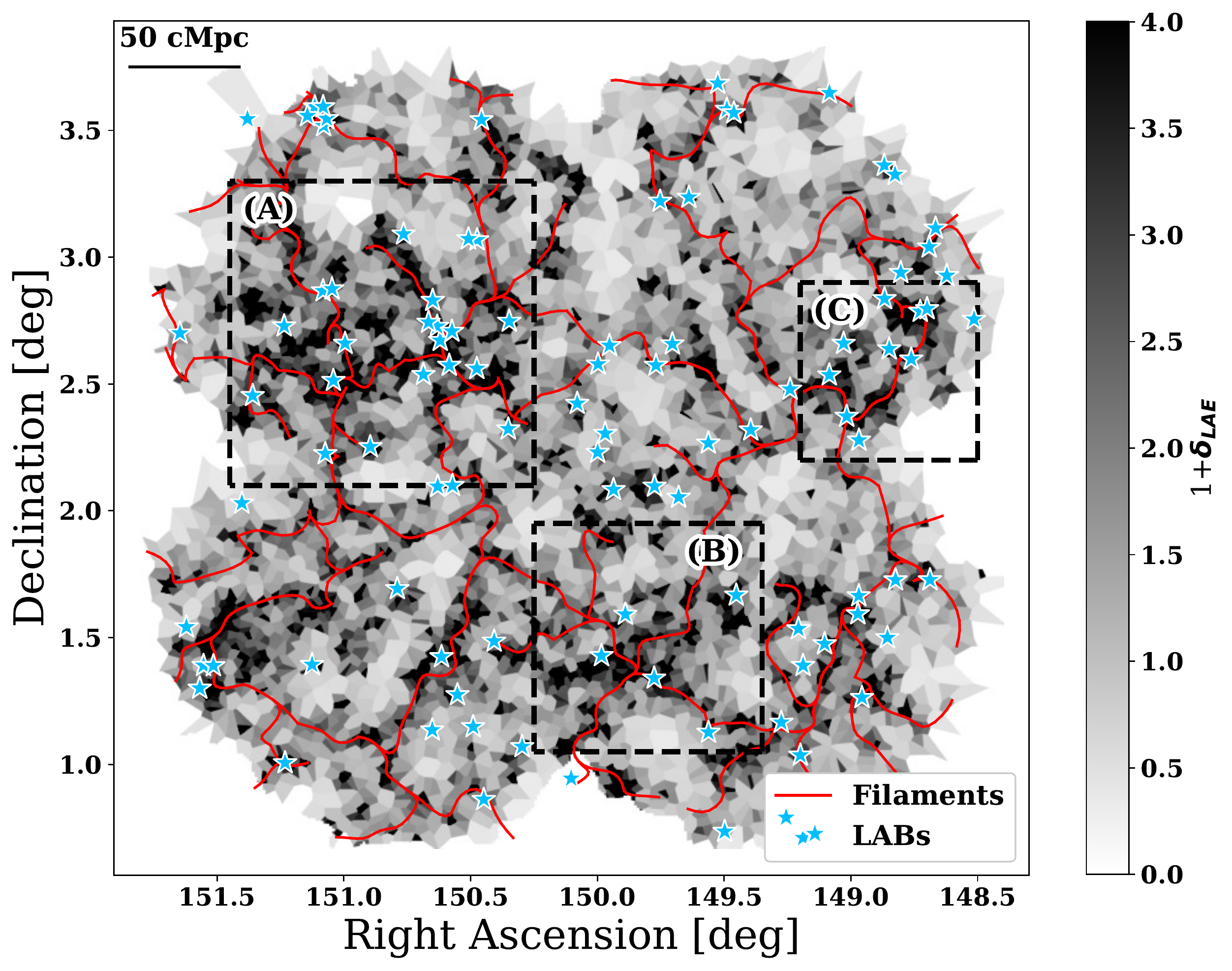}{\linewidth}{}}
    \vspace{-1.5em}
    \gridline{\fig{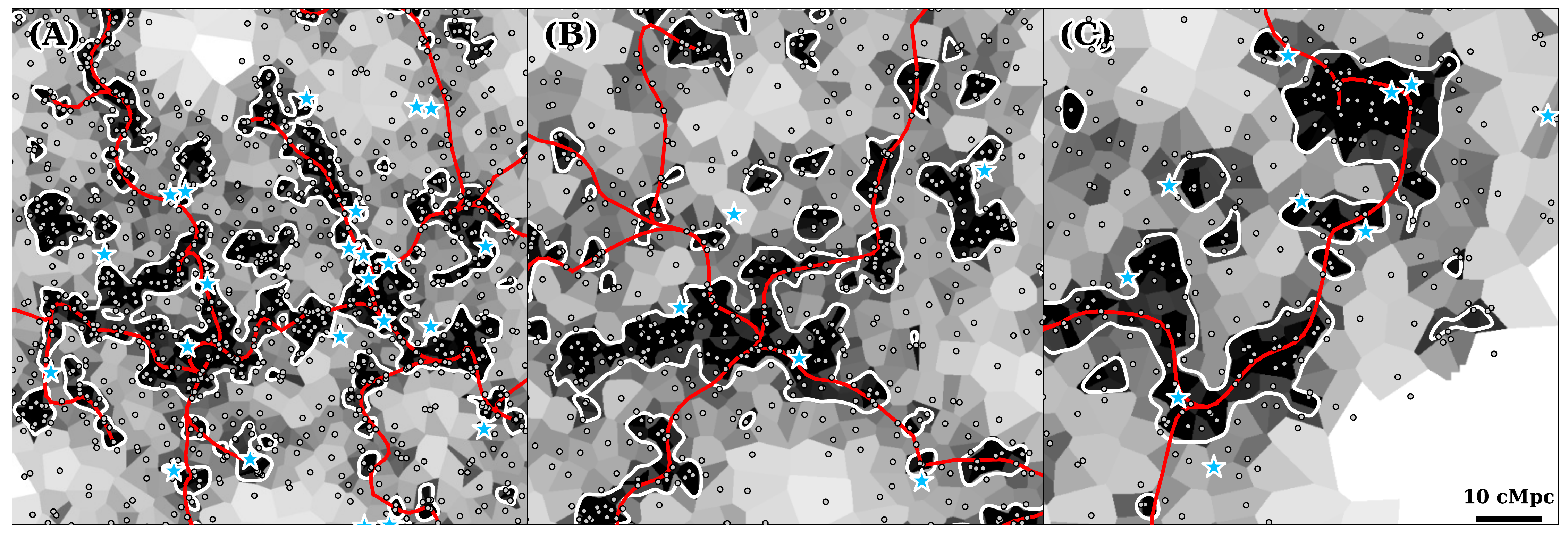}{\linewidth}{}}
    \caption{\emph{Top:} 
    The cosmic filaments traced by LAEs are shown as red lines (see \S~\ref{subsec:filaments} for more detail); overlaid is the VT map indicating LAE densities in greyscale. The bottom panels show the three regions of interest where the white contours outline the 3$\sigma$ overdensity levels. Multiple filaments converge on the most significant structures while adjacent structures are connected by filaments. These configurations are in agreement with the expectations from the hierarchical theory of structure formation.  The locations of LABs (blue stars) relative to the detected filaments strongly hint at the possibility of a close association. 
    }
    \label{fig:filaments}
\end{figure*}

In Figure~\ref{fig:filaments}, we show the cosmic filaments overlaid with the LAB positions and the VT density map. As expected, the filaments generally follow the distribution of LAEs, tracing the intermediate-density regions that connect adjacent overdensity structures. This is illustrated most clearly in the zoom-in views of the Complex A, B, and C. Each structure is connected to multiple filaments, consistent with the expectations from the hierarchical theory. Visual examination  suggests a strong relationship between the positions of LABs and filaments, which will be the subject of our discussion in Section~\ref{subsec:labs_and_filaments}.

\begin{figure*}
    \centering
    \includegraphics[width=6.5in]{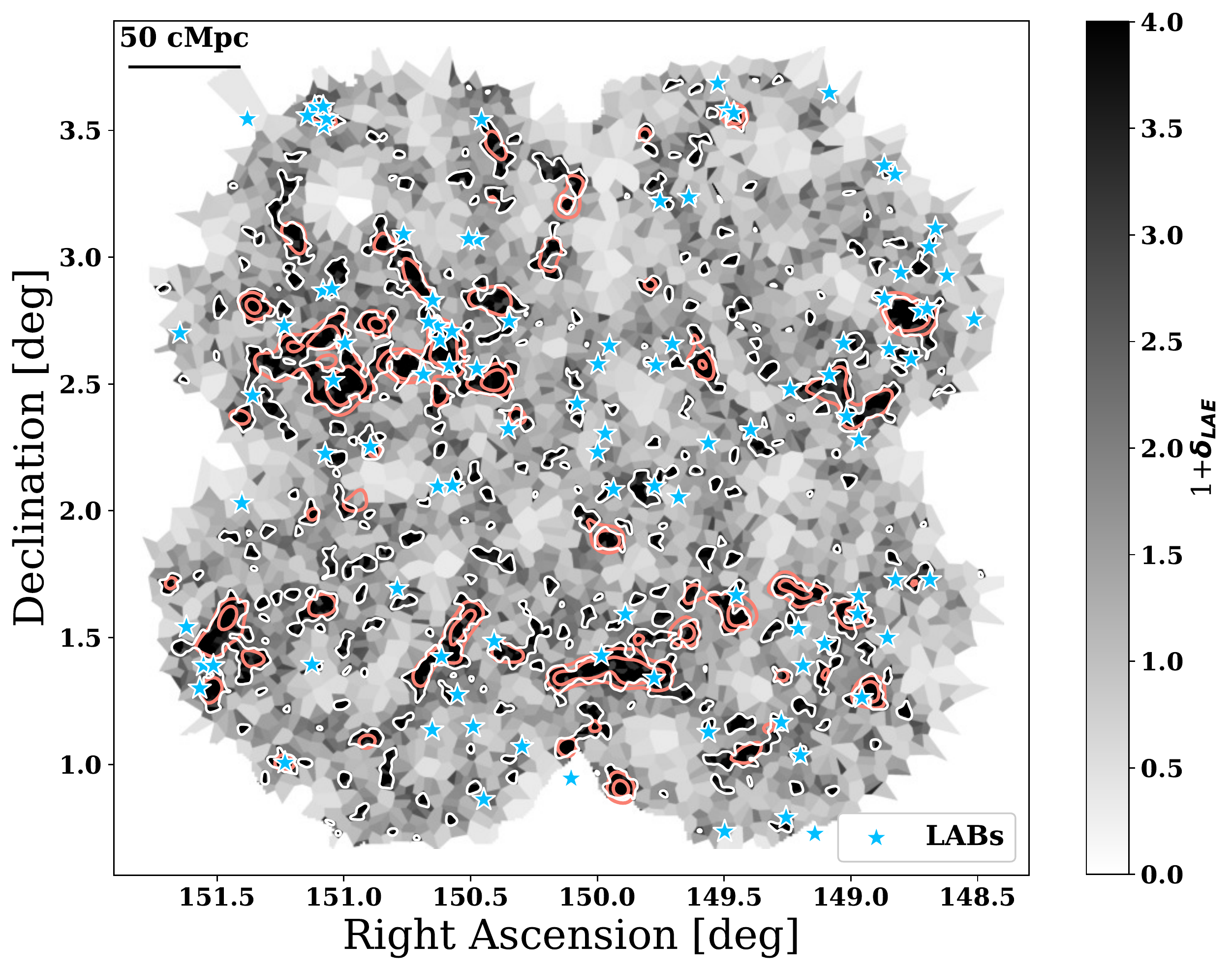}
    \caption{
     Overdensity map of LAEs in the COSMOS field, constructed from the Voronoi tessellation as described in Section~\ref{subsec:voronoi}, with the positions of LABs overplotted. Pink contours indicate 2- and 3$\sigma$ overdensities found in the GS map (Section~\ref{subsec:gauss_smoothing}), while white contours indicate 3$\sigma$ overdensities found in the VT map. It is seen that LABs cluster around overdense structures, and seem to preferentially occupy regions of high LAE overdensity.
     }
    \label{fig:lae_sd_with_labs}
\end{figure*}

\section{The Large-scale Environment of LAB\lowercase{s}} \label{sec:analysis}

\begin{figure*}
    \centering
    \includegraphics[width=5.5in]{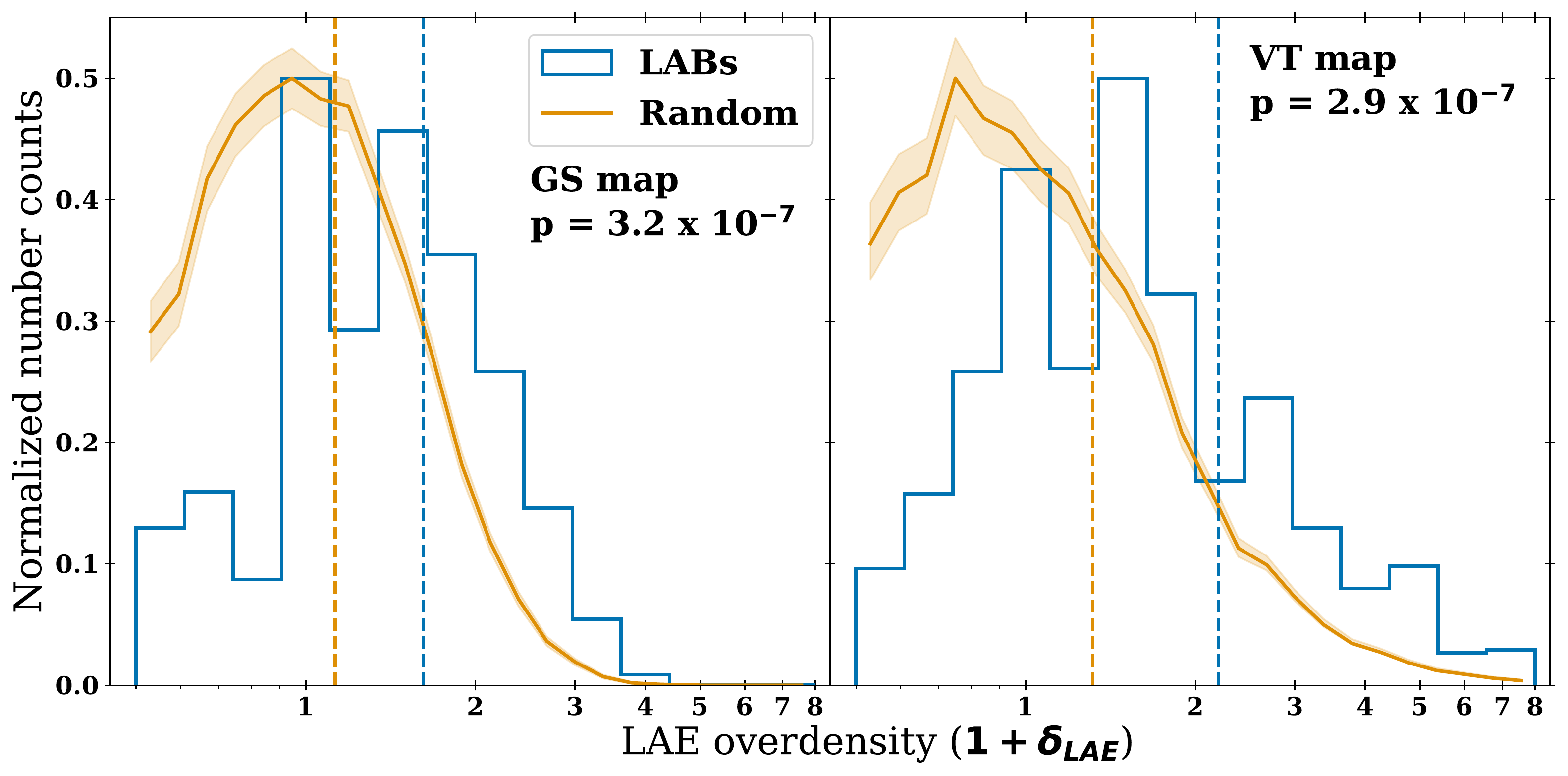}
    \caption{The normalized  distribution of LAE density, \sdrel, at the LAB locations (blue) is compared with that of 5,000 randomly distributed points (orange) as measured from the GS (left) and VT (right) map. The shaded orange region shows the spread of the distribution for different realizations of the random points. {Blue (orange) dotted lines indicate the median value of the distribution for LABs (random points).} The Anderson-Darling test returns the probability of the two samples  being drawn from the same distribution in the order 10$^{-7}$; the $p$ value for each test is indicated at the right bottom corner.} 
    \label{fig:lab_vs_random_sd}
\end{figure*}

Leveraging the indicators of the large-scale structures identified in Section~\ref{sec:lae_ods},  we explore the environment of LABs in this section. Of the 129 LABs, some lie too close to the image boundaries, and as a result lack robust density estimation. While this is the case for both VT and GS maps, the use of a 10~cMpc Gaussian kernel in the GS map additionally leads to the underestimation of the density within $\sim$ 10 cMpc of the edge due to the voids outside it. After removing 27 LABs  for these reasons, we use the sample of 102 LABs from subsequent analyses. 
  
In Figure~\ref{fig:lae_sd_with_labs}, we show the LAB positions overlaid on the VT map where the white contours highlight the $3\sigma$ contours; the $2\sigma$ and $3\sigma$ contours from the GS map are shown in pink. Visual inspection suggests that LABs preferentially reside in regions of moderate to high density. If LABs are randomly distributed, the expectation is that the mode of \sdrel\ at the LAB positions should be 1. Using both GS and VT maps, we find that the mode is \sdrel\ is $\sim$ 1.5 instead, 1$\sigma$ away from that expected for a random spatial distribution. This suggests that LABs are not only clustered but prefer higher-density regions. 

In Figure~\ref{fig:lab_vs_random_sd}, we show the number counts of LABs  and randomly distributed points as a function of \sdrel\  measured in the GS and VT map. Both are normalized to unity. The Anderson-Darling test rejects the possibility that the two samples are drawn from the same underlying distribution at $>$ 99.99\% significance. That LABs populate high-density regions traced by LAEs is extremely unlikely to be due to chance alignment. 

\begin{figure*}
    \centering
    \includegraphics[width=5.5in]{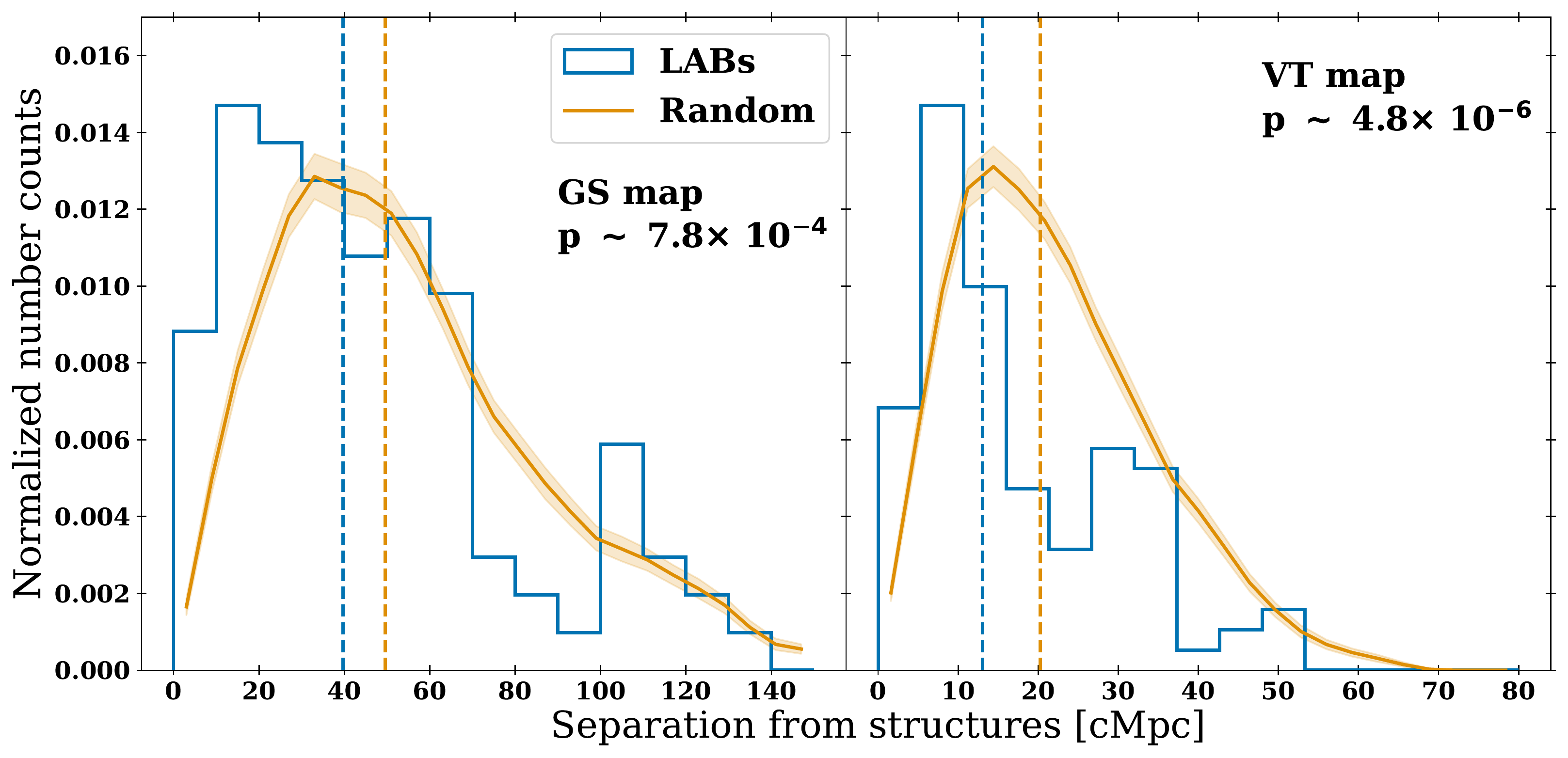}
    \caption{The projected separation of LABs to the nearest overdensity ($d_{\rm LAB,PC}$; blue) is compared with that of random points ($d_{\rm rand,PC}$; orange) where overdensity centers are determined from the GS  (left) and VT map (right). Again the shaded orange region shows the spread of the distribution for different realizations of the random points {and blue (orange) dotted lines indicate the median value of the distribution for LABs (random points).} In both cases, the two distributions are statistically different in that LABs prefer to reside close to galaxy overdensties. 
}
    \label{fig:od_sep}
\end{figure*}

\subsection{Distance of LABs from protoclusters} \label{subsec:pc_sep}

To examine the connection between LABs and overdense structures, we calculate the projected distance of each LAB from the center of the nearest protocluster, which we denote as $d_{\rm LAB,PC}$. Similarly, we populate 5,000 random points within the field and repeat the same measurements ($d_{\rm rand,PC}$). The result, shown in Figure~\ref{fig:od_sep}, suggests that the $d_{\rm LAB,PC}$ distribution peaks at a smaller separation than that of $d_{\rm rand,PC}$, i.e., LABs are located closer to protoclusters than warranted by random distribution.  

The detailed shape of the distribution is sensitive to our definition of {\it a protocluster}. In particular, the separation at which the GS and VT estimates peak is very different. The median value is 39 (48)~cMpc for LABs (random) in the GS map and 13 (20)~cMpc in the VT map. Nevertheless, our results are robust against this variation. The Anderson-Darling test rules out at $>$ 99.999\% significance that LABs and random points are drawn from the same underlying distribution for both the GS and VT maps. While both methods support the hypothesis that LABs prefer to live close to a protocluster, the relative disparity is noteworthy. As discussed in Section~\ref{subsec:finding_struct}, the VT map identifies more and smaller structures than the GS map at the same ($3\sigma$) significance and fares better in detecting and centroiding structures in close separation. Indeed, we find that $\sim$ 26\% of the LABs (27 in number) reside inside a structure identified from the VT map. This could be because the GS map often blends multiple overdensity peaks into one and mislocates the centers thereby washing out the trend. Alternatively, LABs could be associated not only with large protoclusters (easily picked up by the GS method) but also with smaller groups, which the VT method is better at identifying. With the larger LAE samples expected from the ODIN survey, we will be able to disentangle the two possibilities in the near future. 
 
\subsection{LABs and protocluster mass} 

The detection of protoclusters as LAE overdensities allows us to estimate the total mass enclosed therein, which is related to the {\it today mass} of its descendant at $z=0$, $M_{\rm today}$ \cite[e.g.,][]{Steidel2000} provided that the bulk of the mass within the overdensity will fall into the center of the potential well. For each protocluster, we estimate the enclosed mass as follows:
\begin{equation}
\begin{split}
    M_{\rm today} &= \sum_i \rho_{m,i} V_{\rm pix} \\ 
    &= \sum_i \frac{\delta_{{\rm LAE},i}}{b_{\rm LAE}}\rho_0 V_{\rm pix} = \frac{\rho_0(z) V_{\rm pix}}{b_{\rm LAE}}\sum_i \delta_{{\rm LAE},i} 
\end{split}
\end{equation}
where $\rho_{m,i}$ and $\delta_{{\rm LAE},i}$ are the matter density and the LAE overdensity at pixel $i$, respectively; $\rho_0(z)$ is the matter density of the universe at $z=3.1$, $V_{\rm pix}$ is the cosmic volume covered by a single pixel on the VT map, and $b_{\rm LAE}$ is the LAE bias. Each pixel is 120~ckpc on a side covering 0.015 cMpc$^2$ in area. We further assume that the extent of each overdensity is comparable in both line-of-sight and transverse directions, i.e., the cosmic volume spanned by each structure is assumed to be that of a rectangular parallelepiped whose height equals the  square root of the angular area. The LAE bias value is fixed at 1.8 \citep{Gawiser2007}. 

In this simplistic estimate, $M_{\rm today}$ depends sensitively on the definition of a structure -- e.g., the density threshold, and the spatial filter size. In addition, changing the bias value within the range found by existing studies \citep{Ouchi2010,Guaita2010} leads to $\sim$20\% change in $M_{\rm today}$. However, such changes would largely shift the numerical answers for most protoclusters and therefore should not affect any comparative analyses. We plan to evaluate the validity of the assumptions made here by repeating our analyses on the structures in cosmological hydrodynamic simulations (V. Ramakrishnan et al., in prep). 

\begin{figure}
    \centering
    \includegraphics[width=\linewidth]{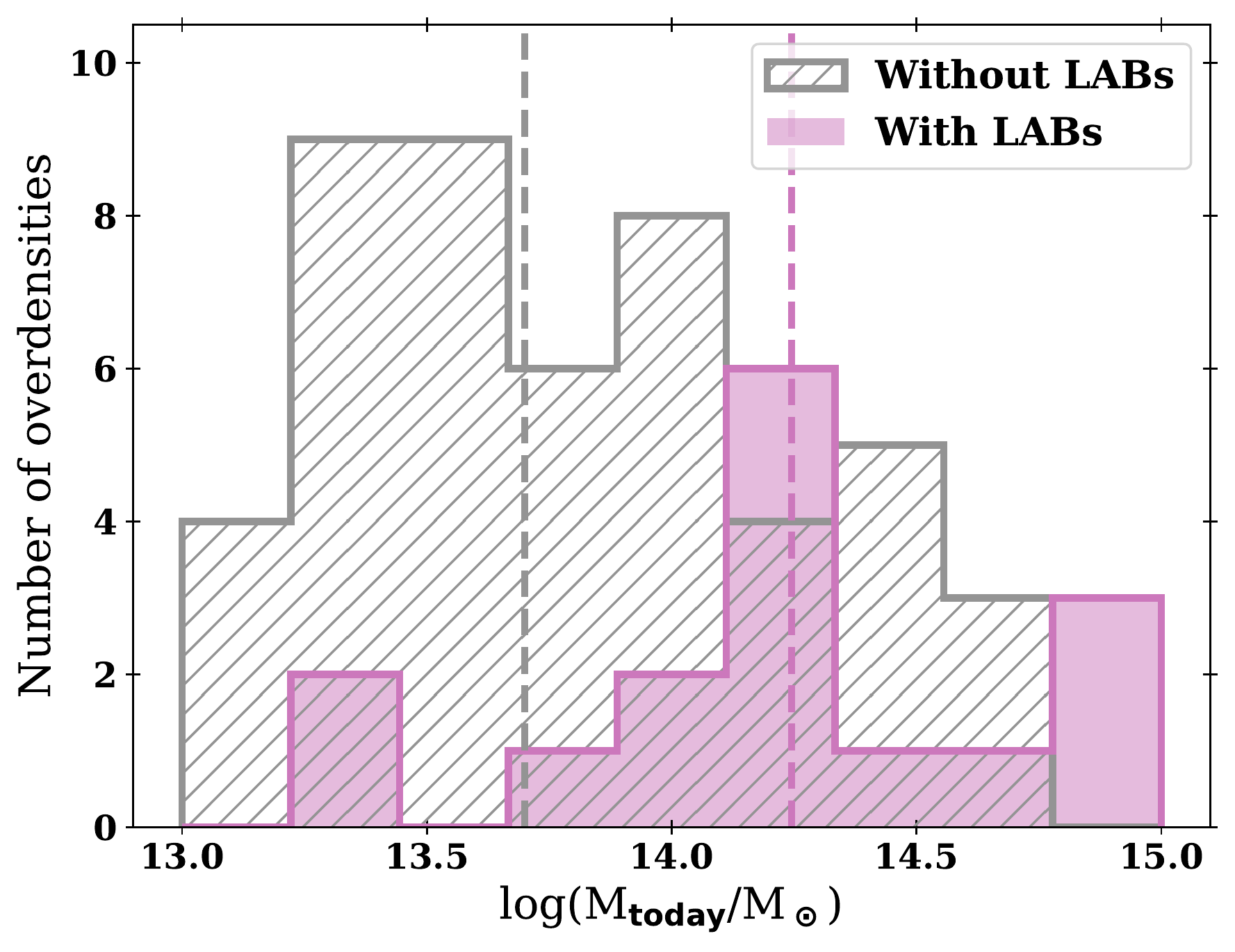}
    \caption{Estimated total masses, $M_{\rm today}$, of protoclusters that host at least one LAB therein (magenta) and those that do not (grey hatched).  The median mass (dashed line) of the former is more massive than that of the latter by a factor of $\approx$3.}
    \label{fig:od_masses}
\end{figure}
In Figure~\ref{fig:od_masses}, we show the $M_{\rm today}$ distributions of the protoclusters which host one or more LABs and of those that do not. Evidently, the two are very different; the two-sample Anderson-Darling test differentiates them at $>$ 99.99\% significance. Protoclusters that host an LAB tend to have much larger $M_{\rm today}$ values than those that do not.  

\begin{figure*}
    \centering
    \includegraphics[width=\textwidth]{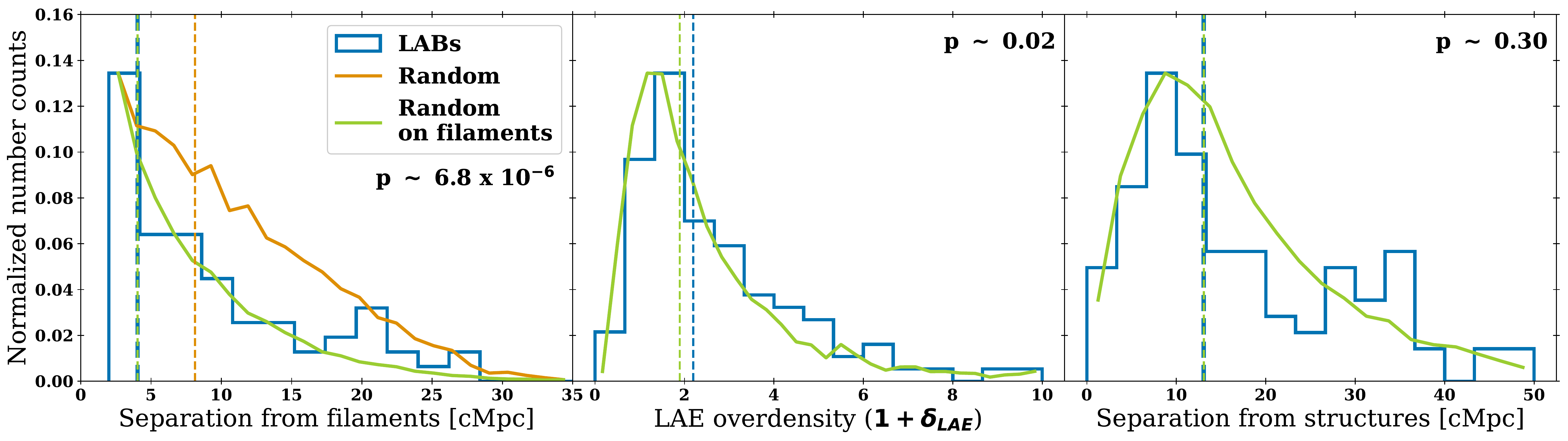}
    \caption{
    The minimum separation from filaments ($d_{fil}$: left), LAE density ($1+\delta_{\rm LAE}$: middle), and minimum separation from protocluster ($d_{\rm PC}$: right) distributions of LABs (blue), random (orange), and random-on-filament points (green).  The $p$ value indicated at the top right corner of each panel is from the two-sample Anderson-Darling test. Filament distance of LABs, $d_{LAB,fil}$, is strongly skewed toward low values relative to a 2D random distribution. When a set of random-on-filaments points that match the $d_{LAB,fil}$ distribution is used as a control sample, the LAB distribution of the minimum distance to protoclusters, $d_{\rm PC}$, is naturally reproduced. The implication is that the primary association of LABs is to filaments and not to protoclusters. 
    }
    \label{fig:lab_sep_fils}
\end{figure*}

\subsection{LABs in the context of cosmic filaments} \label{subsec:labs_and_filaments}

Motivated by the strong correlation between the LABs and cosmic filaments seen in Figure~\ref{fig:filaments}, we explore the physical connection between LABs and the filaments detected by DisPerSE (Section~\ref{subsec:filaments}). A detailed study of the  morphologies of ODIN LAEs in the context of the LSS will be presented in future work.  

We calculate the minimum projected separation of each LAB from the nearest filament, $d_{\rm LAB,fil}$. The same measurements are repeated on a set of 5,000 random points, $d_{\rm rand,fil}$. As shown in the left panel of Figure~\ref{fig:lab_sep_fils}, the two distributions are  different at $>$ 99.99\% confidence. The median projected separation is 4.0 (8.2)~cMpc for the LABs (random). Of the 102 LABs, 77~(75\%) are located within a projected distance of 10~cMpc~(2.4~pMpc) from the nearest filament, and 92~(90\%) are within 20~cMpc~(4.9~pMpc). In comparison, recent hydrodynamic simulations have found that the radius of a filament at $z\sim 3.1$ is  $\sim$2--3~pMpc \citep{Zhu2021}. The significant departure of the $d_{\rm LAB,fil}$ distribution from that of $d_{\rm rand,fil}$ implies that a nonnegligible fraction of the LABs reside inside or close to a filament. Inferring the intrinsic distribution of filament distance from the observed $d_{\rm LAB,fil}$ distribution would require the aid of cosmological simulations  and  realistic modeling of LAEs therein to properly account for the projection effect, which we will present in future work.

Our results in Section~\ref{subsec:pc_sep} show that LABs prefer to live in high-density regions, i.e., near or in protoclusters. Independently, the left panel of Figure~\ref{fig:lab_sep_fils} demonstrates that the same LABs have the propensity to lie close to filaments.  Since filaments are, by definition, ridges of the density distribution that converge at massive (overdense) structures, it is difficult to understand how these two trends are related and if one is causing the other. To disentangle these effects, we create a set of 100,000 points distributed at random along the length of the DiSPerSE filaments while keeping the distribution of their filament separation matched to that of the  LABs.   

In the right panel of Figure~\ref{fig:lab_sep_fils}, we show  the projected separation from protocluster  ($d_{\rm PC}$) of these `{\it random-on-filaments}' points and those of the LABs. The $p$ value returned from the Anderson-Darling test suggests that the two $d_{\rm PC}$ distributions are indistinguishable ($p\sim 0.30$) with similar median values.  The implication is that the preference for LABs to reside near or in cosmic web filaments is the primary driver that leads to their proximity to protoclusters; i.e., the latter trend is simply a byproduct of the fact that large cosmic structures are where many filaments converge. 

However, there exists tentative evidence that filament association may not be the only factor determining where LABs are found. First, LABs are found at slightly higher-density regions than the random-on-filaments points as shown in the middle panel of Figure~\ref{fig:lab_sep_fils}. According to the two-sample Anderson-Darling test, these ($1+\delta_{\rm LAE}$) distributions are different at a $\approx$98\% level. The Komolgorov-Smirnov test returns a consistent result, $p=0.07$, albeit at lower confidence. This is in qualitative agreement with the trend seen in Figure~\ref{fig:od_masses}, that LABs prefer to live in more massive structures. While these trends are not entirely independent of the observed filament association, it opens up a possibility  that LABs may occupy more evolved regions within filaments and/or have a preferred range of density. In future works, we will leverage larger LAB samples and better characterization of filament detection efficiency to fully discriminate different scenarios.

\subsection{Putting it together}

In this work, we have firmly established that LABs \emph{as a population} prefer to occur in overdense regions and in close proximity to protoclusters and cosmic filaments. Our findings are consistent with the fact that some of the known protoclusters host one or more LABs \citep[e.g.,][]{Steidel2000,Matsuda2004,Yang2010,Yang2011,Prescott2012,Saito2015,Caminha2016,Badescu2017,Shi2019}. This could provide some insight into the mechanisms powering LABs - for example, star formation and AGN activity will both be enhanced in overdense regions, which could explain why LABs occur more frequently in these regions. Likewise, in cases where LABs are powered by gravitational cooling \citep[e.g.,][]{Daddi2021}, their luminosity would depend on the host halo mass. Studying the relationship between the size and luminosity of LABs and their environment will be useful in addressing this question, and will be done in a future work.  

The strong association of LABs with cosmic  filaments is not unprecedented. \citet{Umehata2019} detected diffuse Mpc-scale \Lya emission from filaments in the SSA22 protocluster and found that two LABs were embedded within these filaments. They speculated that these LABs were regions of enhanced \Lya emission within the otherwise diffuse and faint gas of the filaments. Our results are  consistent with this picture. \citet{Erb2011} observed that six LABs at $z= 2.3$ form two linear structures spanning $\sim$12 Mpc, along which multiple galaxies at the same redshift lie. Given that the morphology of the LABs also appears to be aligned with these linear structures, they speculated that they trace cosmic filaments \citep[see also][]{Kikuta2019}. 

\citet{Wells2022} found that galaxies in the vicinity of an LAB are on average brighter, more massive and have higher star formation rate than those elsewhere, suggesting accelerated galaxy formation around LABs. The energy from the accelerated galaxy formation could then light up the faint gas of the filaments. In this context, the fact that structures that host LABs are likely to have a higher today mass than those which do not would be due to the fact that a greater number of filaments converge at more massive structures. A useful test would be to see if the morphology of LABs is connected with the nearby filaments as observed in \citet{Erb2011} and \citet{Kikuta2019}. With the large sample size expected from the full ODIN survey, we will be able to robustly quantify  such a relation. 

Finally, several existing studies speculated that LABs are likely associated with group-sized halos \citep[e.g.,][]{Matsuda2006,Yang2010,Yang2011,Prescott2015}. \citet{Badescu2017} reasoned that  blobs  prefer the outskirts of massive structures because they mark the sites of protogroups  accreted onto larger protoclusters. The fact that LABs show some evidence of occupying overdense regions even within filaments (Figure~\ref{fig:lab_sep_fils}, middle panel) may be consistent with this hypothesis. 
With the statistical power afforded by the full ODIN LAB sample, we plan to disentangle the role of filaments, groups, and protoclusters in producing luminous Ly$\alpha$ nebulae, and place more stringent constraints on their formation mechanism. 

\section{Conclusions} \label{sec:summary}

The ODIN survey is currently undertaking deep and wide narrowband imaging of several extragalactic fields totaling $\approx$90~deg$^2$ in area, with the primary aim of identifying Ly$\alpha$-emitting sources at $z=2.4$, 3.1, and 4.5. In this work, we have used the early ODIN science data covering $\sim$ 10 \sqdeg\ in the extended COSMOS field and  identified a sample of 5,352 LAEs and 129 Ly$\alpha$ blobs at $z=3.1$ in the largest contiguous cosmic volume to date spanning $\approx 350 \times 350 \times 70$~(cMpc)$^3$. Using these data, we investigate how LABs are connected to their large-scale environment as traced by LAEs. Our main conclusions are:\\

\noindent{1.} Using the LAE population as a tracer of the underlying matter distribution, we have identified overdense structures as galaxy groups, protoclusters, and filaments of the cosmic web. We find that protoclusters and smaller groups are often strongly clustered together and form extended complexes. The morphologies of these structures are highly irregular and non-spherical; the largest systems are connected to multiple filaments which connect them to smaller structures. These observations are in accordance with expectations from hierarchical structure formation.\\

\noindent{2.} We find that LABs preferentially reside in high-density regions. When compared to randomly located points in the same field, the  $(1+\delta_{LAE})$ distribution of the LABs shows a clear excess and a deficit at the high- and low-density end, respectively.  The two distributions are dissimilar at an extremely high statistical significance, suggesting that our finding is unlikely to be due to chance alignment. \\

\noindent{3.} Starting from the LAE density maps constructed using  Gaussian fixed-kernel smoothing (GS) and Voronoi tessellation (VT), we explore ways to robustly detect cosmic structures, which we broadly refer to as protoclusters. Due to the irregular and often linear/filamentary nature of the angular distribution of the LAEs, we determine that the VT method performs better at detecting protoclusters and at separating them when two are adjacent but distinct. Regardless of the detection method, LABs tend to be located in or near groups and protoclusters with $\approx$30\% of the LABs residing within a structure. Additionally, we find that protoclusters hosting one or more LABs tend to have larger descendant (today) masses than those that do not. \\

\noindent{4.} LABs are also strongly correlated with cosmic filaments. Of our LABs, $\approx$70\% (85) are found within a projected filament distance corresponding to 2.4~pMpc. Given that the radius of a filament at $z=3.1$ is expected to be 2--3~pMpc, our result suggests that a nonnegligible fraction of the LABs reside inside or close to a filament. Inferring the intrinsic distribution of the separation of LABs from cosmic filaments requires the aid of cosmological simulations and realistic modeling of galaxies therein, which we will investigate with the larger samples expected from the ODIN survey. \\

\noindent{5.} The strong association of the LABs to protoclusters and to filaments is connected. When we generate a set of `random-on-filaments' points that match the distribution of projected filament distance of the LABs ($d_{\rm fil}$), the distribution of the minimum separation from protocluster ($d_{\rm PC}$) measured for the LAB sample is naturally reproduced. The implication is that the preference of an LAB to reside near or in cosmic web filaments
is the primary driver that leads to their proximity of protoclusters because large cosmic structures are where many filaments converge.

\begin{acknowledgments}

Based on observations at Cerro Tololo Inter-American Observatory, NSF’s NOIRLab (Prop. ID 2020B-0201; PI: K.-S. Lee), which is managed by the Association of Universities for Research in Astronomy  under a cooperative agreement with the National Science Foundation.
The authors acknowledge financial support from  the National Science Foundation under Grant Nos. AST-2206705 and AST-2206222 and from the Ross-Lynn Purdue Research Foundation Grant. 
BM and YY are supported by the Basic Science Research Program through the National Research Foundation of Korea funded by the Ministry of Science, ICT \& Future Planning (2019R1A2C4069803). 
This work was supported by K-GMT Science Program (GEMINI-KR-2021B-008) of Korea Astronomy and Space Science Institute.
The Institute for Gravitation and the Cosmos is supported by the Eberly College of Science and the Office of the Senior Vice President for Research at the Pennsylvania State University.
AIZ acknowledges support from NSF AST-1715609 and thanks the hospitality of the Columbia Astrophysics Laboratory at Columbia University where some of this work was completed.
MCA acknowledges financial support from the Seal of Excellence @UNIPD 2020 program under the ACROGAL project.
\end{acknowledgments}

\facilities{Blanco (DECam), Gemini:South (GMOS)}

\appendix

\section{Comparing blobs to the general galaxy population}

In Section \ref{sec:analysis}, we demonstrate that LABs live in regions of high LAE density, $(1+\delta_{LAE})$, compared to those of random points. A more pertinent question may be: where are LABs found relative to the general galaxy population? While obtaining a clear answer to this question requires deep wide-field spectroscopy and is thus costly, it  would tell us   more directly about the relationship between LABs and protoclusters and about the preferred range or environmental density or halo mass in which LABs inhabit. 
 Alternatively, we can use LAEs as a representative subset of the underlying galaxy population. 

\begin{figure}
    \centering
    \includegraphics[width=5.5in]{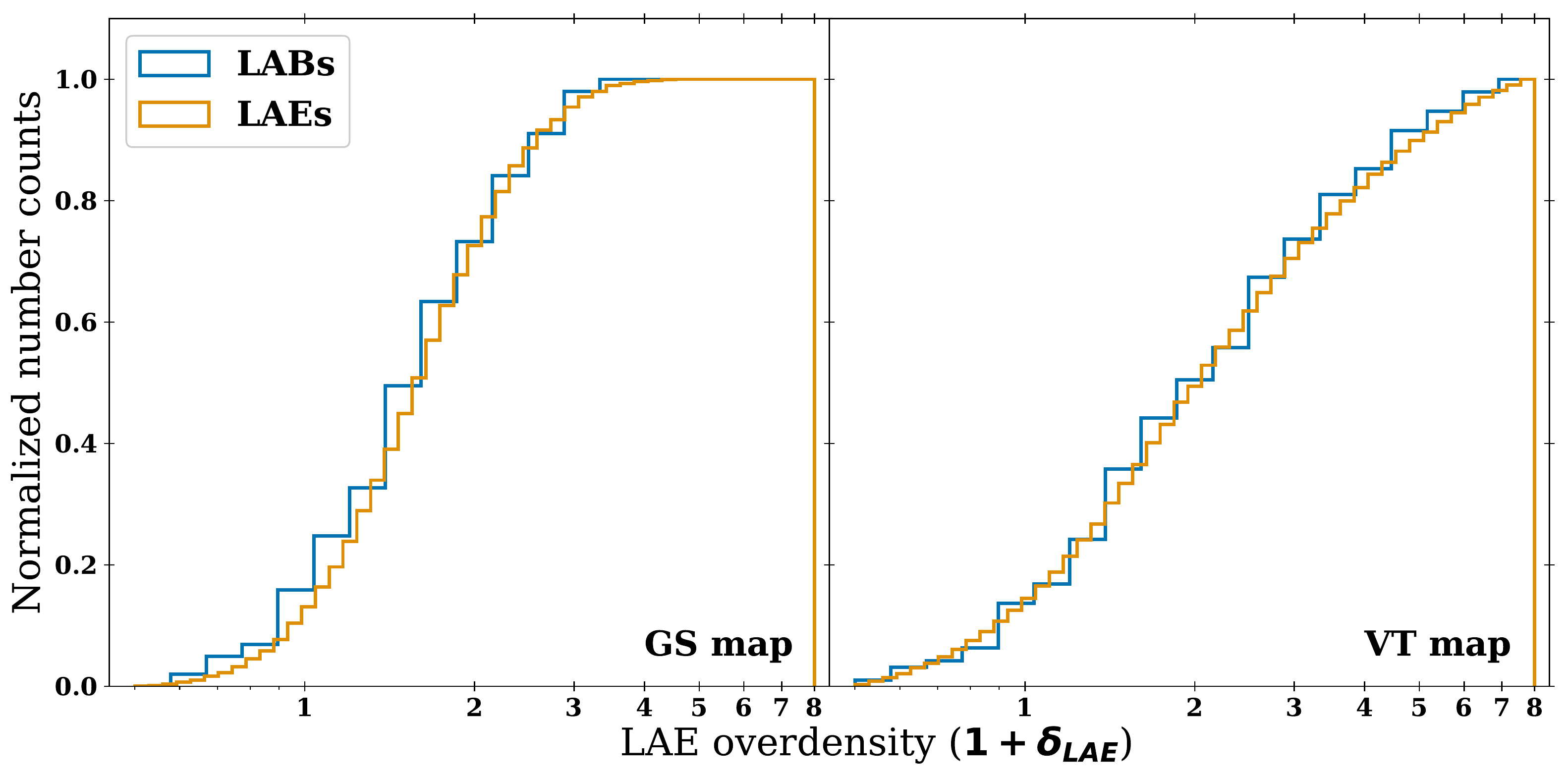}
    \caption{The cumulative distribution of the ($1+\delta_{LAE}$) on the positions of LAEs (orange) and LABs (blue) where the LAE density is measured using the GS (left) and VT (right) method. The two distributions are statistically similar and cannot be distinguished by the Anderson-Darling test.}
    \label{fig:labs_vs_laes}
\end{figure}

In Figure~\ref{fig:labs_vs_laes}, we plot the cumulative $(1+\delta_{LAE})$ distribution of LAEs and LABs. The two are nearly identical as can be seen visually and confirmed by the Anderson-Darling test. Taken at face value, our result suggests that LAEs and LABs intrinsically occupy the same environments; this may seem surprising and even contradictory to our finding that LABs prefer to be near protoclusters and are expected to have a higher galaxy bias. Another possibility is that the two have different distributions but the present data is insufficient to determine it as such. 

We quantify the discriminating power of our dataset by running a test using the IllustrisTNG simulation to address the following question: if LABs reside in more massive halos than LAEs, how well would we be able to detect the trend? To this end, we use the $z=3$ snapshot of the TNG 300~cMpc box and cut out a 60~cMpc slice along the X, Y, or Z direction to match the $N501$ filter width. The transverse size of the TNG300 box is well-matched to our survey field ($7.5~{\rm deg}^2 \approx 9.9\times10^4$~cMpc$^2$ at $z=3.1$ compared to $9.0\times 10^4$~cMpc$^2$ in TNG300). In this volume, we randomly pick dark matter halos above a given mass threshold $M_{min,LAE}$ and assign them as `LAEs'. Similarly, `LABs' are a random subset of the halos above $M_{min, LAB}$, which is set to $10^{12}M_\odot$. The latter assumption is made based on clustering measurements (B. Moon et al., in prep).   The surface densities of these mock LAEs and LABs are matched to those observed in our data.

Using these mock LAE and LAB samples, we repeat the same steps taken in Sections~\ref{subsec:gauss_smoothing} and \ref{subsec:voronoi} and measure the $(1+\delta_{LAE})$ distributions. This procedure is repeated 1,000 times each time reselecting a $300\times 300\times60$~(cMpc)$^3$ subsection of the TNG volume and reassigning LABs and LAEs to a subset of halos therein. In Table~\ref{tab:p_values}, we list the minimum and maximum $p$ values returned by the Anderson-Darling test in these realizations ($p_{\rm min}$ and $p_{\rm max}$) and the fraction in which the two distributions are distinguishable at $>95$\% significance ($f_{p<0.05}$).  We try three $M_{min,LAE}$ values, $10^{9}M_\odot$, $10^{10}M_\odot$, 
 $10^{11}M_\odot$. While we do not vary $M_{min, LAB}$, the expectation is that lower $M_{min, LAB}$ values would mean that LABs have galaxy bias more similar to LAEs, making it more difficult to discriminate the  distributions. The true $M_{min, LAB}$ value  is unlikely to be greater than $10^{12}M_\odot$. If all halos with masses $M\geq 10^{12}M_\odot$ host an LAB, the LAB surface density would be comparable to that observed in our data. 
From the table, only in 17\%--22\% of the realizations are the two distributions meaningfully different regardless of the minimum halo mass assigned to LAEs. In light of this, it is not surprising that we are unable to distinguish the two distributions in the real data given the current sample size. We take a step further and forecast how well our measurements will improve once the full ODIN data at $z=3.1$ is at hand, which will be nine times larger than the current dataset. 
%
%
The result is shown in the bottom half of Table~\ref{tab:p_values}. The fraction in which the two distributions are distinguishable, $f_{p<0.05}$, is significantly higher at 74--98\%.  However, the range of $p$ values remains wide, implying that it is not a guaranteed outcome. 

The observed large statistical uncertainty is in part due to small number statistics for LABs combined with the high cosmic variance expected for massive halos hosting them. In addition, we remind readers of another caveat. Since LAEs themselves are used to compute the overdensity, the high end of the $(1+\delta_{\rm LAE})$ distribution for LAEs is bound to be overrepresented compared to any other galaxy sample. For example, if one `pixel' in the density map contains six LAEs therein, we would count them six times instead of one. A more equitable comparison may be made using galaxy samples identified regardless of their Ly$\alpha$ emission, e.g., stellar-mass or $M_{\rm UV}$ limited sample. With $\approx$1,000 LABs at each redshift ($z=2.4$, 3.1, and 4.5) expected at the completion of the ODIN survey, measurement of angular clustering to infer their host halo masses remains a viable alternative.


\begin{deluxetable}{cccccccccc}
\tablecaption{\label{tab:p_values}}
\tablecolumns{10}
\tablehead{ \colhead{} & \multicolumn{3}{c}{$M_{min,LAE}$ = 10$^9$ M$_\odot$} & \multicolumn{3}{c}{$M_{min,LAE}$ = 10$^{10}$ M$_\odot$} & \multicolumn{3}{c}{$M_{min,LAE}$ = 10$^{11}$ M$_\odot$} }
     \startdata
     & $p_{\rm min}$ & $p_{\rm max}$ & $f_{p < 0.05}$ & $p_{\rm min}$ & $p_{\rm max}$ & $f_{p < 0.05}$ & $p_{\rm min}$ & $p_{\rm max}$ & $f_{p < 0.05}$ \\
     \hline
    \multicolumn{10}{c}{{\bf ODIN COSMOS}} \\
    \hline
     GS map & 6.6 $\times$ 10$^{-5}$ & 1.0 & 0.17 & 8.4 $\times$ 10$^{-5}$ & 1.0 & 0.16 & 1.2 $\times$ 10$^{-5}$ & 1.0 & 0.22 \\
     \hline
     VT map & 3.4 $\times$ 10$^{-5}$ & 1.0 & 0.17 & 3.3 $\times$ 10$^{-5}$ & 1.0 & 0.17 & 1.7 $\times$ 10$^{-5}$ & 1.0 & 0.21 \\
     \hline\hline
    \multicolumn{10}{c}{{\bf ODIN Full }} \\
    \hline
     GS map & 5.6 $\times$ 10$^{-7}$ & 0.77 & 0.81 & 5.6 $\times$ 10$^{-6}$ & 0.88 & 0.79 & 2.9 $\times$ 10$^{-7}$ & 0.36 & 0.98 \\
     \hline
     VT map & 1.8 $\times$ 10$^{-6}$ & 0.77 & 0.61 & 1.3 $\times$ 10$^{-5}$ & 0.93 & 0.60 & 2.2 $\times$ 10$^{-6}$ & 0.80 & 0.74 \\
     \enddata
     \tablecomments{Comparison of the $(1+\delta_{LAE})$ distributions of mock `LAEs' and `LABs' selected in the Illustris TNG using the Anderson-Darling test. Mock LAEs are assigned to a random subset of halos with masses $M_{min,LAE}$. We fix $M_{min,LAB}$ to $10^{12}M_\odot$. The minimum/maximum $p$ values and the fraction in which the two distributions are different at $>95$\% level ($f_{p<0.05}$) are based on 1,000 realizations.
     }
\end{deluxetable}


\section{Filaments of the cosmic web with different detection settings}\label{appendix_B}
\begin{figure*}
    \centering
    \includegraphics[width=\linewidth]{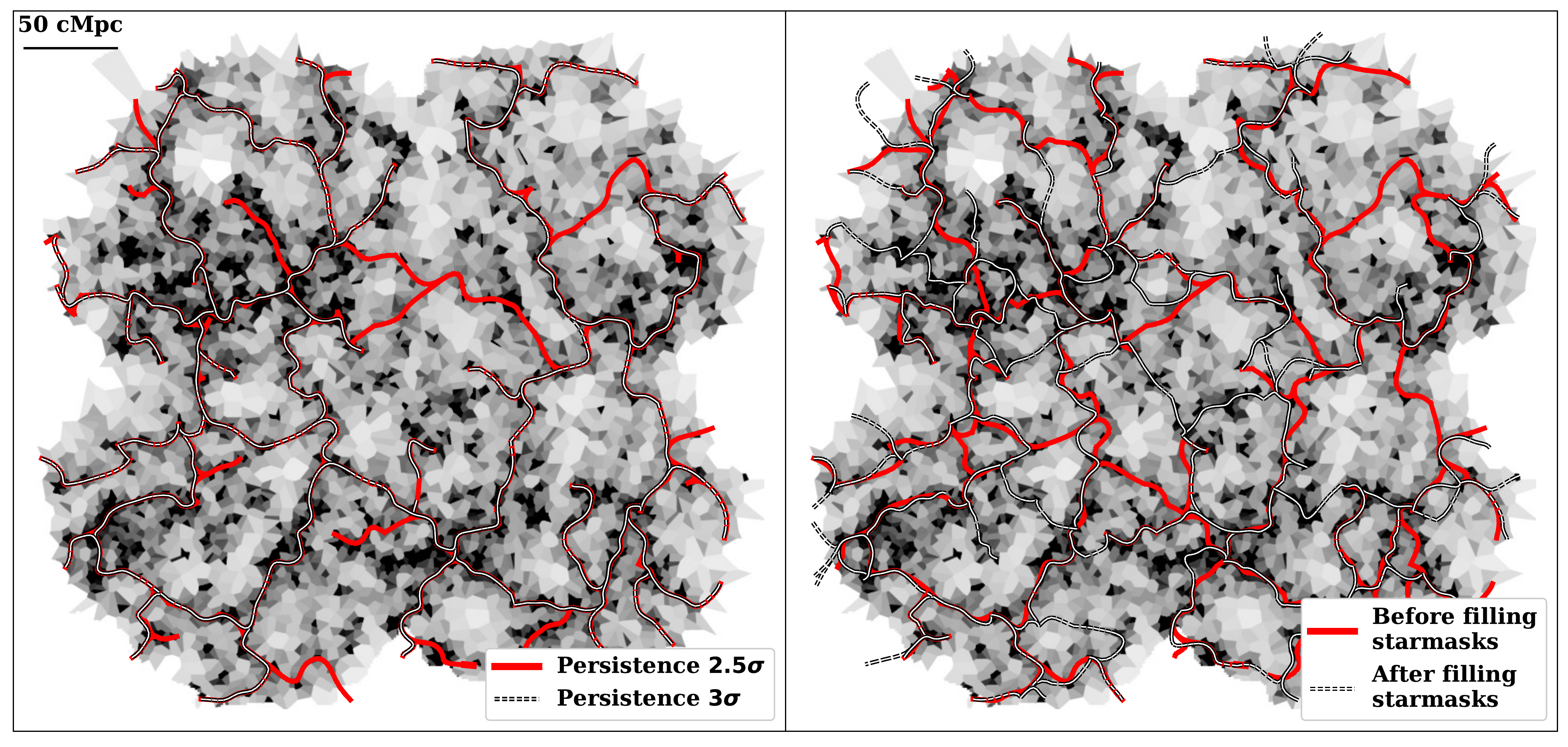}
    \caption{DisPerSE-detected filaments are shown as red and white segments when we change the persistence level (left) and fill in the regions of bright stars with random points (right). The background in both panels is the VT-based $(1+\delta_{LAE}$) density map. While the number of filaments depends on the run setting, a majority of filaments in and around the densest regions are detected in both. Our main conclusions remain robust against these changes.}
    \label{fig:fil_settings}
\end{figure*}

In running DisPerSE, we use the {\tt -btype smooth} option, which generates additional points outside the field boundaries via interpolation intended to mitigate the edge effect. The choice of persistence is important as a higher persistence setting extracts more robust but less detailed filamentary structures. We set persistence to 2.5$\sigma$, slightly lower than those used in recent studies \citep[e.g.,][]{Kraljic2017,Malavasi2016}. In the left panel of Figure~\ref{fig:fil_settings}, we show the filaments identified using persistence set to $2.5\sigma$ and $3\sigma$. While most of the filaments in the regions of interest (e.g., Complex A, B, and C) are detected with both persistence values, one long structure in Complex A is undetected when persistence is set to $3\sigma$. Since the same region also shows an excess of LABs (see the bottom panel (A) in  Figure~\ref{fig:filaments}), we set it to $2.5\sigma$ for our final set of filaments. 

We also test how the regions excluded by bright star masks  affect our ability to meaningfully identify filaments. This is done by filling in the masked regions with a random set of points commensurate with the field LAE density. The result is shown in the right panel of Figure~\ref{fig:fil_settings}. While the two sets of filaments are not identical, only the shortest filaments tend to be significantly affected. The majority of the filaments are left unchanged. Using either set of filaments does not change our main conclusions.

\bibliography{refs}{}
\bibliographystyle{aasjournal}

\end{document}